\documentclass[aps,prb,reprint,superscriptaddress]{revtex4-1}
\usepackage{times}
\usepackage{graphicx}
\usepackage{amsmath}
\usepackage{color}

\begin{document}


\title{Microwave nanobolometer based on proximity Josephson junctions}


\author{J. Govenius}
\email[]{joonas.govenius@aalto.fi}

\author{R. E. Lake}

\author{K. Y. Tan}

\author{V. Pietil\"{a}}
\affiliation{QCD Labs, COMP Centre of Excellence, Department of Applied Physics, Aalto University, P.O. Box 13500, FIN-00076 Aalto, Finland}

\author{J. K. Julin}

\author{I. J. Maasilta}
\affiliation{Nanoscience Center, Department of Physics, P.O. Box 35, FIN-40014 University of Jyv\"{a}skyl\"{a}, Finland}

\author{P. Virtanen}
\affiliation{O.~V.~Lounasmaa Laboratory, Aalto University, P.O. Box 15100, FIN-00076 Aalto, Finland}


\author{M. M\"{o}tt\"{o}nen}
\affiliation{QCD Labs, COMP Centre of Excellence, Department of Applied Physics, Aalto University, P.O. Box 13500, FIN-00076 Aalto, Finland}
\affiliation{O.~V.~Lounasmaa Laboratory, Aalto University, P.O. Box 15100, FIN-00076 Aalto, Finland}


\date{\today}
\begin{abstract}
We introduce a microwave bolometer aimed at high-quantum-efficiency
detection of wave packet energy within the framework of
circuit quantum electrodynamics,
the ultimate goal being single microwave photon detection.
We measure the differential thermal conductance
between the detector and its heat bath, obtaining values
as low as $5 \mbox{ fW/K}$ at $50 \mbox{ mK}$.
This is one tenth of
the thermal conductance quantum and corresponds to a theoretical lower bound on
noise-equivalent-power of order $10^{-20} \mbox{ W/}\sqrt{\mbox{Hz}}$ at $50 \mbox{ mK}$.
By measuring the differential thermal conductance of the same bolometer design in
substantially different environments and materials, we determine
that electron--photon coupling dominates the thermalization of our nanobolometer.
\end{abstract}


\maketitle 


\section{Introduction}

Reaching noise-equivalent-power (NEP) of
$10^{-20} \mbox{ W/}\sqrt{\mbox{Hz}}$
in radiation sensors is an important
goal for space-based telescopy
because it allows cosmic background radiation limited
spectroscopy in the THz regime.\cite{karasik-IEEETHZ-2011, benford-nimpr-2004}
What intrigues us however, is that such low noise levels
would also enable direct measurement of wave packet energy
in circuit quantum electrodynamics (cQED).\cite{blais-pra-2004, wallraff-nature-2004}
For example, NEP of $10^{-21} \mbox{ W/}\sqrt{\mbox{Hz}}$ allows
resolving an individual $10 \mbox{ GHz}$ microwave photon
emitted from a qubit or resonator with energy relaxation time of $10 \mbox{ } \mu \mbox{s}$.
Even a more modest NEP of $10^{-18} \mbox{ W/}\sqrt{\mbox{Hz}}$
enables on-chip phase-insensitive energy measurements of multi-photon wave packets, such as
dispersive qubit measurement pulses.\cite{blais-pra-2004,riste-nature-2013}
Since the energy of a wave packet and the voltage produced by it
correspond to non-commuting quantum-mechanical observables,
such single-shot energy measurements are fundamentally limited in precision
when using traditional microwave amplifiers which amplify voltage.\cite{nation-rmp-2012}
The difference between these measurement schemes is crucial in single-shot measurements
of non-classical pulses containing a definite amount of energy,
such as in the Hong-Ou-Mandel effect or in linear optics quantum computation.\cite{hong-prl-1987,knill-nature-2001}
Although the Hong-Ou-Mandel effect can be verified by ensemble averaging traditional voltage measurements,\cite{dasilva-pra-2010,lang-nphys-2013}
that approach is not scalable to more complex experiments that
require feedback conditioned on energy measurements.
In addition to low NEP, such feedback experiments require high bandwidth
as well as absorption and detection of nearly all of the incident microwave radiation, i.e., high quantum efficiency.
On the other hand, bolometry in cQED does not require a broad dynamic range
or antenna coupling to off-chip radiation sources.

Transition edge sensors\cite{wei-natnano-2008, karasik-apl-2011, karasik-IEEETHZ-2011, kenyon-spie-2006, santavicca-apl-2010, karasik-apl-2012} (TES)
and kinetic inductance detectors\cite{day-nature-2003, janssen-apl-2013, devisser-natcom-2014}
are the most mature low-temperature bolometer technologies.
Use of 
other superconductor weak links,\cite{nahum-physicab-1994, schmidt-apl-2003, schmidt-apl-2005, prober-ieee-2007}
semiconductor nanostructures,\cite{komiyama-nature-2000, hashiba-nanotech-2010, komiyama-ieee-2011}
graphene,\cite{fong-prx-2012, yan-natnano-2012, mittendorff-apl-2013, mckitterick-jap-2013}
carbon nanotubes,\cite{kawano-apl-2009}
and quantum capacitance\cite{richards-apl-1980, bueno-apl-2010, echternach-apl-2013}
has also been experimentally explored.
State-of-the-art nanoscale TESs,\cite{karasik-apl-2011} semiconducting detectors,\cite{komiyama-ieee-2011}
and quantum capacitance detectors\cite{echternach-apl-2013}
have reached phenomenal NEP
at THz frequencies
but their quantum efficiency relies on high-energy input photons compared to
a device-specific energy scale, e.g.,
the minimum energy for breaking Cooper pairs in a quantum capacitance detector
or the energy above which the TES impedance is well approximated
by its normal-state resistance.\cite{sadleir-prl-2010}
Furthermore, TESs and semiconducting detectors are typically
read out using low-bandwidth amplifiers.\cite{wei-natnano-2008, karasik-apl-2011, komiyama-ieee-2011}
Our hot-electron\cite{wellstood-prb-1994} nanobolometer addresses the
low-frequency impedance matching issue by including
a nanoscale resistive absorber element
that is thermally strongly coupled to the thermometer element,
but in a configuration that allows independent electrical design and operation of the two elements.
This allows absorbing all incoming radiation down to arbitrarily low frequency
by matching the resistance of the absorber element
to the characteristic impedance of the input transmission line, typically $50\mbox{ }\Omega$ in cQED.
The thermometer element on the other hand is mostly reactive,
enabling the use of a fast rf-coupled readout technique similar to quantum capacitance detectors.
Probing changes in a reactive rather than a resistive thermometer
allows the use of a larger readout power for a given maximum
tolerable level of measurement-induced heating.
This is important for minimizing the effect of noise added by
the rest of the amplification and digitization circuitry
and hence for approaching the theoretical limits on NEP.
We also note that our detector
demonstrates experimentally the temperature to inductance transduction mechanism
proposed in the so called Josephson proximity sensor.\cite{giazotto-apl-2008, voutilainen-jap-2010}

We report on measurements of thermal conductance between our nanobolometer
and its heat bath.
This is an important first step toward demonstrating feasibility of our design
since thermal conductance $G$ is an essential parameter in determining the magnitude of thermal energy fluctuations
between the bolometer and its heat bath.
These fluctuations set a lower bound on NEP because temperature measurements
cannot distinguish them from variations in input signal power.\cite{mather-ao-1982,moseley-jap-1984,booth-arnps-1996}
Generally, fluctuations of order $\sqrt{ (G T) (k_B T) }$ arise from shot noise
intrinsic to any Poisson process that transports an average power $G T$
in packets of typical size $k_B T$,
but the exact expression depends on details of the thermalization
and thermoelectric feedback mechanisms.\cite{mather-ao-1982}
We find that for our rf coupled sensors the differential
thermal conductance is $5 \mbox{ fW/K}$ at $50 \mbox{ mK}$.
This implies that the theoretical
lower bound on NEP set by thermal energy fluctuations
is of order $10^{-20} \mbox{ W/}\sqrt{\mbox{Hz}}$.
We note that similar thermal conductances have been previously achieved in suspended
TESs\cite{kenyon-spie-2006} and even lower values in a
hot-electron TES,\cite{wei-natnano-2008}
but without impedance matching at microwave input frequencies
and with substantially lower readout bandwidth.

By measuring the same bolometer design in different electromagnetic environments
and by using different materials, we find that the dominant heat link between
our sensor and the environment is the electron--photon thermal conductance.
Like other single-mode conduction channels, the electron--photon conductance
is bound from above by the universal quantum of thermal conductance\cite{pendry-1983, schwab-nature-2000, schmidt-prl-2004, meschke-nature-2006, jezouin-science-2013}
$G_{Q} = \pi^2 k_b^{2} T / 3h$,
which is reached when the detector and environment impedances
are matched at thermal excitation frequencies.\cite{pascal-prb-2011}
By engineering the electromagnetic environment in the vicinity of the sensor,
we reduce the total differential thermal conductance to
one tenth of $G_{Q}$ at $50 \mbox{ mK}$, i.e., $5 \mbox{ fW/K}$.
This value is likely dominated by parasitic electromagnetic coupling to the environment.

\section{Principle of operation}

\begin{figure}
\includegraphics[width=8.6cm,keepaspectratio]{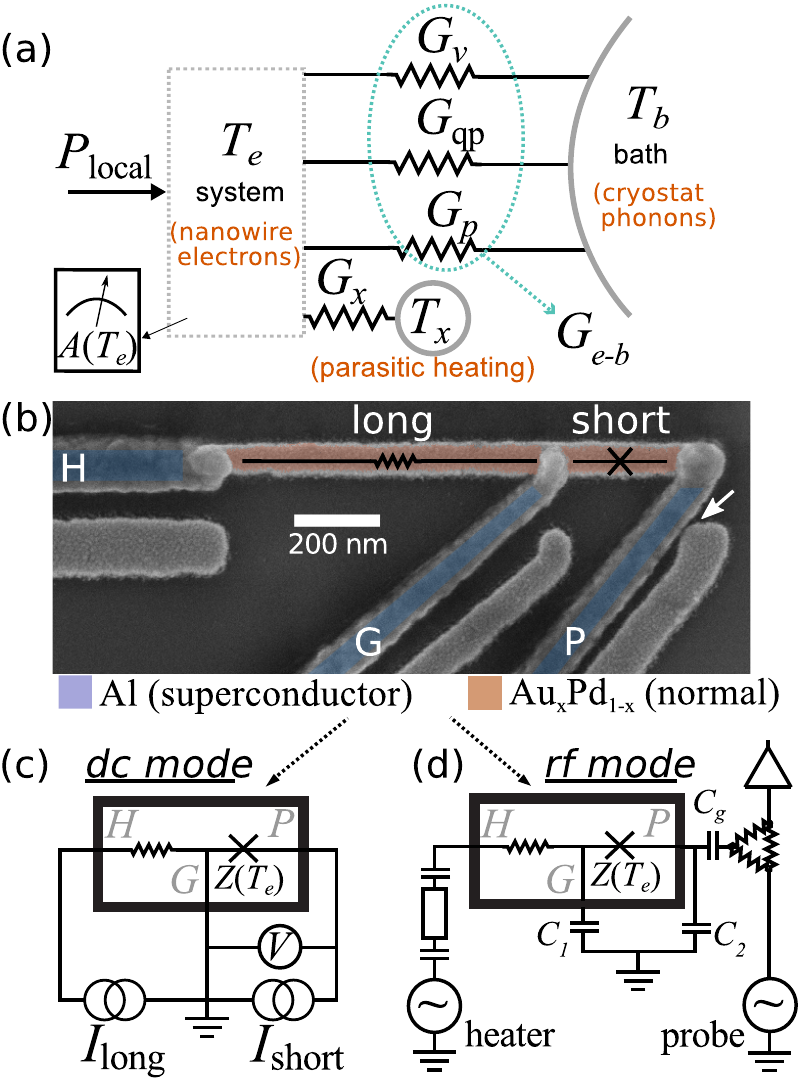} 
\caption{(Color online)
(a) Thermal model of the electron system at temperature $T_{e}$ coupled to a bath at $T_{b}$
through photonic ($G_{\nu}$), quasiparticle ($G_{\textrm{qp}}$),
and phononic ($G_{p}$) thermal conductances
that comprise the total conductance ($G_{e-b}$).
Here, $G_{x}$ quantifies coupling to a parasitic bath ($T_{x}$),
$P_{\textrm{local}}$ is local heating power, and $A(T_{e})$ is a temperature-dependent observable.
(b) Micrograph of Sample \textsc{A}.
The normal-metal nanowire is contacted by three superconductor leads:
heater~(H), ground~(G) and temperature probe~(P).
The white arrow indicates an unintentional galvanic contact present only in Sample \textsc{A}.
(c) Simplified electrical schematic for measurement of the
temperature-dependent sensor impedance $Z(T_{e})$
in Samples \textsc{A}, \textsc{B}, \textsc{C}, and \textsc{N} (dc mode).
(d) Same as (c) but for Samples \textsc{R} and \textsc{F} (rf mode).
\label{fig:setup}}%
\end{figure}

The central component of our nanobolometer is a diffusive normal-metal nanowire
contacted by three superconductor leads,
which together form two diffusive superconductor--normal-metal--superconductor (SNS) junctions
with a total normal-metal volume of order $(100 \mbox{ nm})^{3}$
[see Fig.~\ref{fig:setup}(b) and Table~\ref{tab:params}].
We fabricated all nanowires with electron beam lithography using
shadow angle evaporation of
$\textrm{Au}_{x}\textrm{Pd}_{1-x}$ as normal metal
and Al or Nb as superconductor on oxidized silicon substrates.
We estimate $x \approx 0.75$ based on evaporation parameters and known alloy stabilities.\cite{okamoto-1985}
Electrical measurements were performed in a cryostat with base temperature of 10 mK.
At a bath temperature of 100 mK, the long junction
is an ohmic resistor $R_{N,\textrm{long}}$ in all samples,
while the short junction supports
a non-dissipative supercurrent with clear switching and retrapping at
currents of order 100 nA.
Even at 10 mK, the long junction shows
a mere reduced resistance at the smallest currents ($<10 \mbox{ nA}$)
but no switching.
This allows us to neglect the reactive component of the long junction admittance.

Samples \textsc{A}, \textsc{B}, \textsc{C}, and \textsc{N} were dc coupled and measured using sub-kHz frequencies [dc mode, Fig.~\ref{fig:setup}(c)],
while Samples \textsc{R} and \textsc{F} were capacitively coupled and measured at microwave frequencies [rf mode, Fig.~\ref{fig:setup}(d)].
Shunting lead G [Fig.~\ref{fig:setup}(b)] to ground
through a small resistance in dc mode
and by a large capacitor ($C_{1}$) in rf mode
prevents electrical cross-talk between
the long and short junctions during device operation,
but does not prevent thermalization\cite{peltonen-prl-2010, bezuglyi-prl-2003} between
the junctions on relevant timescales.
Therefore, a single temperature $T_{e}$ accurately describes
the electronic system of the entire normal-metal nanowire.
In all samples, the long junction heats the electron gas ($P_{\textrm{local}}$)
while the short junction transduces $T_{e}$ into an electrical signal.
In dc mode, a fixed current bias heats the electrons with 
$P_{\textrm{local}} = I^{2}_{\textrm{long}}R_{N,\textrm{long}}$,
while the current at which the short junction switches from
the superconducting state to the normal state indicates electron temperature,
a technique known as proximity-effect thermometry.\cite{dubos-prb-2001,giazotto-rmp-2006,meschke-jltp-2009}
In rf mode, a transmission line delivers $P_{\textrm{local}}$ at several GHz,
while a small amplitude excitation at hundreds of MHz probes the $T_{e}$
dependent inductance of the short junction.

\subsection{dc mode}

In dc mode, we ramp the current bias through
the short junction $I_{\textrm{short}}(t)$
linearly over several
milliseconds while measuring voltage $V$
over it in a four-wire configuration.
As the short SNS junction switches from the
superconducting state to the normal state,
$V$ jumps from zero to $I_{\textrm{short}}R_{\textrm{N,short}}$ at a stochastic time $t_{\textrm{switch}}$.
Since the ramp is much slower than the inverse bandwidth of our electrical lines,
we can convert $t_{\textrm{switch}}$ directly to a current $I_{\textrm{short}}(t_{\textrm{switch}})$.
We repeat this $I_{\textrm{short}}$ bias cycle approximately $10^3$ times and
record a histogram of switching currents [Fig.~\ref{fig:data}(a)]
as well as the voltage trace averaged over all repetitions.
We then define either the median of the switching current distribution
or the current at which $V$ on average crosses $0.5 \times I_{\textrm{short}}R_{\textrm{N,short}}$
as a typical switching current $I^{\star}_{s}$.
This $I^{\star}_{s}$ defines an uncalibrated electron temperature probe $A(T_{e})$ [Fig.~\ref{fig:setup}(a)] in dc mode.
We call it uncalibrated because we do not attempt to deduce
the absolute temperature $T_{e}$ or the theoretical critical current $I_c$
from $A(T_{e})$.
Note also that the heating current $I_{\textrm{long}} \ll I^{\star}_{s}$.

A key assumption in our model is that $I^{\star}_{s}$ does not vary with $T_{b}$ at constant $T_{e}$.
Mathematically, we ignore $|\partial_{T_b} I^{\star}_{s}(T_e,T_b)|$
as negligible compared to $|\partial_{T_e} I^{\star}_{s}|$
within the temperature range of interest.
Physically, we assume that phase fluctuations across the junction
are damped by the electron gas at $T_{e}$,
rather than the external environment at $T_{b}$.
This is reasonable for SNS junctions given their high plasma frequency and
strong dissipation at high frequencies.\cite{virtanen-prb-2011}
Furthermore, the similarity of switching current distributions
along contours of $I^{\star}_{s}$ [see Fig.~\ref{fig:data}(a)]
supports the validity of the assumption;
it suggests that the distributions are well described by a single parameter,
which would be surprising if $|\partial_{T_b} I^{\star}_{s}| \sim |\partial_{T_e} I^{\star}_{s}|$.
Also note that overdamped electrical response at high frequencies
is not contradictory with the observed hysteretic switching and retrapping behavior
which can arise due to Joule heating.\cite{courtois-prl-2008}


\subsection{rf mode}

For Sample \textsc{R} (\textsc{F}), the long junction absorbs $P_{\textrm{local}}$
from a monochromatic 8.8~GHz (6.74~GHz) coherent excitation applied between leads H and G [Fig.~\ref{fig:setup}(d)].
The heating tone is generated by a room-temperature microwave generator
and delivered to the sample through fifty-ohm coaxial transmission lines, a number of
commercial attenuators and filters inside the cryostat,
and finally a
symmetrically coupled
on-chip co-planar waveguide (CPW) resonator
with a fundamental resonance frequency equal to the heater frequency.
We calibrated the attenuation of the commercial components,
assumed negligible attenuation for resonant
transmission through the overcoupled CPW resonator,\cite{goppl-jap-2008}
and estimated $R_{\textrm{N,long}}$ based on resistivities of the dc coupled samples
in order to take into account the small (${\sim}15\%$)
amount of power reflected due to $R_{\textrm{N,long}} > 50 \, \Omega$.
The CPW resonator is strongly coupled to the transmission lines,
which leads to a low loaded quality factor of $10^2$ compared
to typical internal quality factors of $10^4$, hence justifying
the full transmission assumption.
The resonator acts as a Lorentzian bandpass filter that
isolates the detector from non-thermal noise at other frequencies
and reduces the electron--photon thermal conductance,
which would otherwise be close to $G_Q$ due to the intentional
matching between the characteristic impedance of
the heating line and the long junction.
We note that similar band or low pass filtering is in general
practical in cQED where the thermal frequency $k_B T_b / h$
is typically much smaller than the photon frequency of interest.

We probe the electron temperature
through the reflection coefficient of a tank circuit
that consists of on-chip parallel plate capacitors $C_{1} \sim 100 \mbox{ pF}$, $C_{2}$, and $C_{g}$
together with the mostly reactive admittance $Z(T_{e})^{-1}$
of the short SNS junction [Fig.~\ref{fig:setup}(d)].
For Sample \textsc{R} (\textsc{F}), $C_{1}=25C_{2}=75C_{g}$ ($C_{1}=C_{2}=10C_{g}$).
In the samples measured in rf mode, the short junction in fact consists of six SNS junctions made of alternating
$150-200 \mbox{ nm}$ pieces of superconductor and normal-metal,
but they are treated as one effective admittance in this article.
In linear response, the tank circuit in Sample \textsc{R} (\textsc{F})
is a harmonic oscillator
with a resonance frequency
$f_{0} \approx 1 / \big( 2 \pi \sqrt{LC} \big)$
of $1.3 \mbox{ GHz}$ ($430 \mbox{ MHz}$) and quality factor of ten (hundred).
Here $C^{-1} = C_{1}^{-1} + C_{2}^{-1}$
and $L = -1/(\omega \operatorname{Im} \{Z(T_{e})^{-1}\}) \sim 1 \mbox{ nH}$
is the effective junction inductance at angular frequency $\omega$.
The quality factor is governed by the external coupling capacitor $C_{g}$
and internal losses within the resonator.\cite{dassonneville-prl-2013, virtanen-prb-2011}
As the junction heats up, $L \propto 1/I_{c}$
increases and hence $f_{0}$ decreases.
We define $f_{0}$ as $A(T_{e})$ in rf mode and extract it by
measuring the reflection over a range of frequencies near $f_{0}$.
The measurement signal is generated and digitized at room temperature,
but the sample is protected from high-temperature noise 
by a number of attenuators and amplifiers inside the cryostat.
We separate the input and output of the reflection measurement with a
resistive splitter and isolate the sample from amplifier noise with attenuators.
These should be replaced by circulators in future bolometric applications requiring high signal-to-noise ratio.

As in dc mode, we assume that $f_{0}$ does not vary with $T_{b}$ at constant $T_{e}$.
In addition to the reasons discussed in the previous section,
this assumption is justified by the nearly linear response of the SNS junction at low currents.
Nonlinearities that couple incoherent fluctuations to the reflection measurement
become important only when the current fluctuations become comparable to $I_c$,
i.e., only on the energy scale of the Josephson energy $E_J > 2 \mbox{ K}$.
The nearly linear behavior together with the small heating currents
($\sqrt{ P_{\textrm{local}} / R_{\textrm{N,long}}} \sim \mbox{nA}$)
also prevents cross-talk between the heating and thermometry signals,
even in Sample \textsc{F} where $C_{1}=C_{2}$.

\section{Differential thermal conductance}

We define the differential thermal conductance
between the nanowire electrons ($e$) and cryostat phonons ($b$) as
the increase in the power flow between the two as $T_b$ is decreased
and $T_{e}$ is kept constant, i.e.,
\begin{equation}
\tilde{G}_{e-b} = -\frac{\partial P_{e-b}(T_{e},T_{b})}{\partial T_{b}} \mbox{,}
\end{equation}
where $P_{e-b} = (T_{e} - T_{b})G_{e-b}$ is the net power flow from $e$ to $b$
and $G_{e-b}(T_{e},T_{b})$ is the non-differential thermal conductance [Fig.~\ref{fig:setup}(a)].
Our definition is closely related to previous definitions of differential\cite{mather-ao-1982}
or dynamic\cite{richards-jap-1994} thermal conductance,
but with the roles of $T_{e}$ and $T_{b}$ exchanged.
In steady state, the temperatures satisfy the power balance equation
\begin{equation} \label{eq:pwr-balance}
P_{\textrm{local}} + P_x(T_{e},T_{x}) = P_{e-b}(T_{e},T_{b}) \mbox{,}
\end{equation}
where a $T_b$-independent power $P_x$ quantifies unintentional parasitic heating,
which we represent as a weak coupling $G_{x}$ to an independent temperature bath ($T_{x}$)
in Fig.~\ref{fig:setup}(a).
The parasitic heating may be due to radiation
leaking in from warmer parts of the cryostat
or due to non-thermal noise from electronics.
Because $\tilde{G}_{e-b}$ is independent of this parasitic heating, we can
compare it directly to the theoretical predictions for different channels,
such as $\tilde{G}_{e-b} = \pi^2 k_b^{2} T_b / 3h$ for matched single-channel conduction
or $\tilde{G}_{e-b} = 5 \Sigma V_0 T_{b}^{4}$ for electrons thermalized
by bulk phonons in a metal of volume $V_0$ and coupling strength $\Sigma$.
Note that $T_{e}$ appears in these predictions
only for non-linear channels that include cross terms of $T_{b}$ and $T_{e}$ in $P_{e-b}$.
By measuring $\tilde{G}_{e-b}$, we can therefore
experimentally investigate which thermalization channels dominate,
even without a calibrated $T_{e}$ sensor.
Furthermore, $\tilde{G}_{e-b}$ is the correct quantity for estimating
the thermal-energy-fluctuation limited NEP due to coupling to $T_{b}$,
which we consider to be of greater interest than
fluctuations in the parasitic heating
originating from imperfect shielding and filtering.

\subsection{Isothermal technique and measured $\tilde{G}_{e-b}$}

Our method of obtaining $\tilde{G}_{e-b}$ is based on mapping
contours of constant $A(T_{e})$ in the $(T_{b}, P_{\textrm{local}})$-plane.
Specifically, $-\tilde{G}_{e-b}$ is given by the slope
of a contour of $A(T_{e})$ in the $(T_{b}, P_{\textrm{local}})$-plane,
as seen by differentiating Eq.~(\ref{eq:pwr-balance}) with respect to $T_{b}$
while holding $T_{e}$ constant.
This is the same principle as in the so called isothermal technique
used previously in TES-type samples,\cite{gershenzon-sjltp-1988, elantev-sjltp-1989, karasik-ieee-2009}
except that we take parasitic heating into account
and therefore need to distinguish $\tilde{G}_{e-b}$ from $G_{e-b}$
even in the $P_{\textrm{local}} \rightarrow 0$ limit.
Appendix~\ref{app:thermal} discusses some caveats of the isothermal technique
that may arise if a more general thermal model than that of Fig.~\ref{fig:setup}(a) is required to capture the physics of the system.

\begin{figure}
 \includegraphics[width=8.6cm,keepaspectratio]{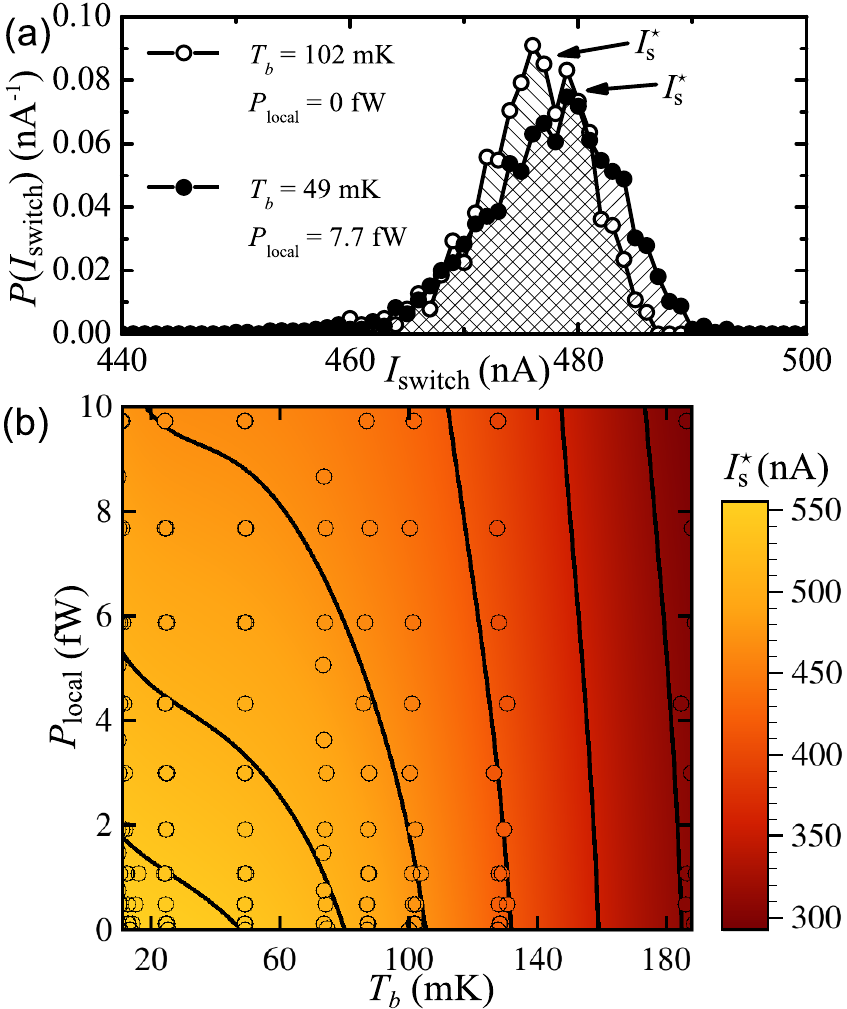}
 \caption{(Color online) (a) Switching probability density as a function of current through the short SNS
   junction. The arrows indicate the extracted typical switching currents $I_{s}^{\star}$.
   (b) Measured $I_{s}^{\star}$ (color inside circles) at different
   bath temperatures $T_b$ and heating powers $P_{\textrm{local}}$.
   Black curves indicate contours of constant $A(T_{e})=I_{s}^{\star}$,
   as determined from a smoothing function
   (color outside circles). Both panels are for Sample \textsc{A} (see Table~\ref{tab:params}).
   \label{fig:data}}%
 \end{figure}

We apply this method to the measured typical switching currents
shown in Figure \ref{fig:data}(b) for Sample \textsc{A}.
In order to draw smooth contours of $A(T_{e})$,
we fit a phenomenologically chosen smooth function
$I_{s}^{\star}(T_{b}, P_{\textrm{local}})$ to the data points;
the details of the smoothing function do not affect the
extracted $\tilde{G}_{e-b}$ averaged over a $T_{b}$-scale larger than the data point spacing.
We then compute $-\partial P_{\textrm{local}} / \partial T_{b} = \tilde{G}_{e-b}$ in the $P_{\textrm{local}} \rightarrow 0$ limit for
many such contours and show the resulting curve in Fig.~\ref{fig:g},
along with the results from all other samples.
The results are similar for gradients computed at non-zero $P_{\textrm{local}}$ (not shown).

The temperature range in Fig.~\ref{fig:g} is fundamentally limited
at the high end by the breakdown of the assumption that
the heating current is negligible compared to the thermometer current.
At the low end, the vanishing magnitude of $\partial_{T_b}A(T_{e}(T_b,P_{\textrm{local}},P_x))$
prevents extracting $-\partial P_{\textrm{local}} / \partial T_{b}$ mainly
due to slow drifts in $P_x$ on the time scale of several hours.
In practice, a limited amount of data also restricts the $T_b$ range
of some of the curves, e.g., for Sample \textsc{C}.

\begingroup
\squeezetable
\begin{ruledtabular}
\begin{table*}[htbp]
  \centering
  \caption{Parameters for the long and short sections [Fig.~\ref{fig:setup}(b)]
    of the nanowire in each sample. Here, $t$ is the nanowire thickness measured by a quartz crystal
    deposition monitor. We extracted lengths and widths from micrographs and measured the normal-state
    resistance $R_{N}$ at $T_{b}=10\mbox{ mK}$ in dc mode. In rf mode, $R_{N}$ is an estimate based on size and typical resistivity. We also provide the experimentally observed differential thermal conductance $\tilde{G}_{e-b}$ at 70~mK bath temperature.
  }
    \begin{tabular}{@{}lll|llll|llll|lll@{}}
           &        &&  &\textit{Long section}     &       &         &    &   \textit{Short section} &       &        &$T_{b} = 70$ mK &  \\
    sample & t (nm) & & length (nm) & width (nm) & $R_{N}$ $(\Omega)$ & &   length (nm) & width (nm) & $R_{N}$ $(\Omega)$ & & $\tilde{G}_{e-b}$ (fW/K) &  \\\colrule
    A & 20    && 720   & 80    & 120 &&   250   & 100     & 40  &  & 130   &  \\
    B & 20    && 980   & 150   & 80  &&   310   & 140    & 30  &  & 40    &  \\
    C & 20    && 750   & 50    & 300 &&   273   & 60     & 100 &  & 30    &  \\
    N(Nb) & 60    && 1700  & 440   & 6.8 &&  180    & 440    & 4.3 &  & 80   &  \\
    R & 25    && 1400  & 140   & 100 (est.) &&  6 $\times$ 180 & 200   &     &  & 10    &  \\
    F & 20    && 1400  & 130   & 130 (est.) &&  6 $\times$ 180 & 140   &     &  & 10    &  \\
    \end{tabular}%
  \label{tab:params}%
\end{table*}%
\end{ruledtabular}
\endgroup

\subsection{Comparison to predicted mechanisms}

\begin{figure}[t]
\includegraphics[width=8.6cm,keepaspectratio]{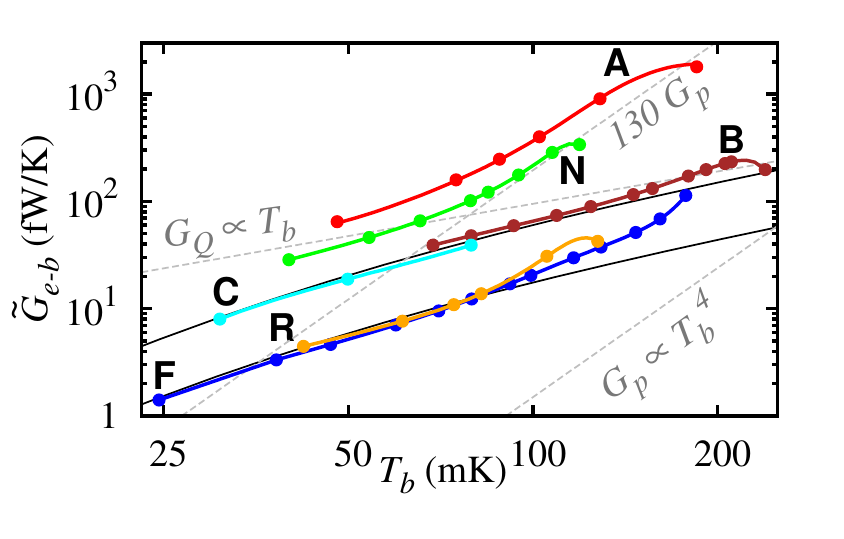}
\caption{(Color online) Differential thermal conductance $\tilde{G}_{e-b}(T_b)$ (solid curves with points)
  for dc samples (\textsc{A}, \textsc{B}, \textsc{C}, and \textsc{N})
  and rf samples (\textsc{R} and \textsc{F}) against bath temperature $T_b$.
  Each point indicates the $T_b$ value for a set of measured data [see Fig.~\ref{fig:data}(b)].
  The dashed lines show the quantum of thermal conductance $G_{Q}$
  and the phonon contribution $G_{p}$ for a typical nanowire volume of $(100 \mbox{ nm})^3$
  and for a large volume of $130 \times (100 \mbox{ nm})^3$.
  Solid black curves correspond to the contribution of electron--photon coupling
  given a phenomenological model described in Appendix~\ref{app:emenv} and Fig.~\ref{fig:emmodel}.
  \label{fig:g}}
\end{figure}

We analyze Fig.~\ref{fig:g} in terms of contributions
arising from the nanowire electrons being coupled to substrate phonons,\cite{wellstood-prb-1994}
photons in the electromagnetic environment,\cite{pascal-prb-2011, meschke-nature-2006}
and quasiparticles in the superconductor
leads.\cite{muhonen-rpp-2012,chandrasekhar-sst-2009}
We do not expect a measurable contribution from phase slips in the leads.\cite{arutyunov-2008}

Coupling of the nanowire electrons to substrate phonons
is expected to contribute $5 \Sigma V_0 T_{b}^4 = G_{p}$ to $\tilde{G}_{e-b}$,
where $\Sigma \approx (3 \pm 1) \times 10^{9} \mbox{ W/m$^{3}$K$^{5}$}$ for $\textrm{Au}_{.75}\textrm{Pd}_{.25}$\cite{timofeev-prl-2009b, muhonen-apl-2009} and $V_0=(100 \mbox{ nm})^{3}$
is a typical volume of our nanowires.
This leads to the estimate $G_{p} = 2 \mbox{ fW/K}$ at
$T_{b} = 100 \mbox{ mK}$, which is one to two orders of magnitude less than the measured values.
Furthermore, the observed scaling of $\tilde{G}_{e-b}$ is much weaker than $T_{b}^4$
at the lowest temperatures (Fig.~\ref{fig:g}),
suggesting that the phonon channel is insignificant for
Samples \textsc{B}, \textsc{C}, \textsc{R}, and \textsc{F}.
Above 80~mK, Samples \textsc{A} and \textsc{N} on the other hand approach the
expected dependence for thermalization via phonons in a
volume of roughly $130 \times (100 \mbox{ nm})^{3}$.
This is similar to the measured volume of Sample \textsc{N} [$70 \times (100 \mbox{ nm})^{3}$]
and can be explained for Sample \textsc{A} by an accidental
galvanic contact between one of the Al leads
and the tip of the corresponding normal-metal shadow [see Fig.~\ref{fig:setup}(b)].
In that case, thermal resistance over the short superconducting link
between the nanowire and the shadow is negligible\cite{peltonen-prl-2010} and
thermalization is limited by electron-phonon coupling in the combined
nanowire--shadow volume of approximately $130 \times (100 \mbox{ nm})^{3}$,
leading to results comparable to Courtois et al.\cite{courtois-prl-2008}

In the quasiparticle channel,
electrons with energy higher than the superconductor energy gap $\Delta$ diffuse
into the three superconductor leads.\cite{chandrasekhar-sst-2009}
The thermal conductance for long superconductor leads is suppressed from the normal state value by
a factor of $6(y^{2}+2y+2)e^{-y} / \pi^{2}$
where $y = \Delta/(k_{B}T_{\textrm{qp}})$.\cite{peltonen-prl-2010, bezuglyi-prl-2003, bardeen-pr-1959}
At quasiparticle temperature $T_{\textrm{qp}}=100 \mbox{ mK}$ this corresponds to only a few aW/K even for Al.
However, the effective $T_{\textrm{qp}}$ can be much higher than $T_{b}$,\cite{aumentado-prl-2004, devisser-prl-2011, muhonen-rpp-2012}
so we chose to increase $\Delta$ in Sample \textsc{N} by using Nb instead of Al.
Since at low temperatures $\tilde{G}_{e-b}$ for Sample \textsc{N} is similar to the other dc mode samples (Fig.~\ref{fig:g}),
we conclude that quasiparticles in the superconductor leads
do not significantly add to the total heat conductance.
This argument also applies to quasiparticles excited by multiple Andreev reflections.\cite{bezuglyi-prb-2001}

Electron--photon coupling contributes a volume and material independent
term to $\tilde{G}_{e-b}$, with a theoretical maximum value of $2G_{Q} \propto T_b$
which is reached when both junctions are perfectly matched to a resistive environment.
In general, the matching varies as a function of frequency,
causing deviations from the linear $T_b$ dependence.
We indeed observe a $T_b$ dependence that falls between linear and quadratic
in the low-temperature regime (except for Sample \textsc{A}).
This is consistent with thermalization through
poor but non-vanishing matching of the sample and
environment impedances over a broad range of thermally excited frequencies.
Furthermore, $\tilde{G}_{e-b} / G_{Q}$ is notably smaller
for rf samples (${\sim}0.1$) than dc samples (${\sim}0.5$)
which differ essentially only in their coupling
to the electromagnetic environment. Thus we attribute the observed low-temperature heat conductance to the photonic channel.

Since we did not intentionally engineer strong electromagnetic coupling
to the sample, the exact physical structure constituting
the dissipative environment remains unknown.
Instead, in Appendix~\ref{app:emenv} we use a simple model of parasitic
capacitive coupling to a resistive environment. This model reproduces the observed low-temperature thermal conductances shown in Fig.~\ref{fig:g} with realistic values of the free parameters (see Fig.~\ref{fig:emmodel}).

\section{Conclusion}

We introduced a hot-electron nanobolometer
integrated with
a microwave transmission line input
and a broad-band rf readout.
We measured the differential thermal conductance
for multiple samples in substantially different materials
and electromagnetic environments
and found values as low as $5 \mbox{ fW/K}$ at $50 \mbox{ mK}$,
attributed to parasitic electron--photon coupling.
This value implies a fundamental
thermal-energy-fluctuation limited noise
level that is low enough for
applications of great practical interest,
in particular on-chip
bolometric measurements in
circuit quantum electrodynamics.
Demonstrating such low noise-equivalent-powers in practice
will require technical improvements in the amplification of
the readout signal and possibly in sample shielding,
if fluctuations in parasitic heating end up limiting performance.
Finally, we introduced a precise definition of differential
thermal conductance in the presence of a
parasitic heating term,
which is non-negligible in many experiments at millikelvin temperatures.

\begin{acknowledgments}
We thank Tero Heikkil\"{a} for helpful discussions
and Leif Gr\"{o}nberg for depositing Nb thin films on some samples.
We also gratefully acknowledge the provision of facilities and technical support by Aalto
University at Micronova Nanofabrication Centre
and the financial support from
Emil Aaltonen Foundation,
the European Research Council under Grant 278117 (``SINGLEOUT''),
Academy of Finland under Grants 135794, 138903, 141015, 260880, 265675, 272806, and 251748 (``COMP''),
the Finnish Cultural Foundation,
and the European Metrology Research Programme (``EXL03 MICROPHOTON'').
The EMRP is jointly funded by the EMRP participating countries within EURAMET and the European Union.
We also thank all developers of the open-source lab environment \textsc{qtlab}.
\end{acknowledgments}

\appendix

\section{Extended thermal models}\label{app:thermal}

\begin{figure}[t]
\includegraphics[width=8.6cm,keepaspectratio]{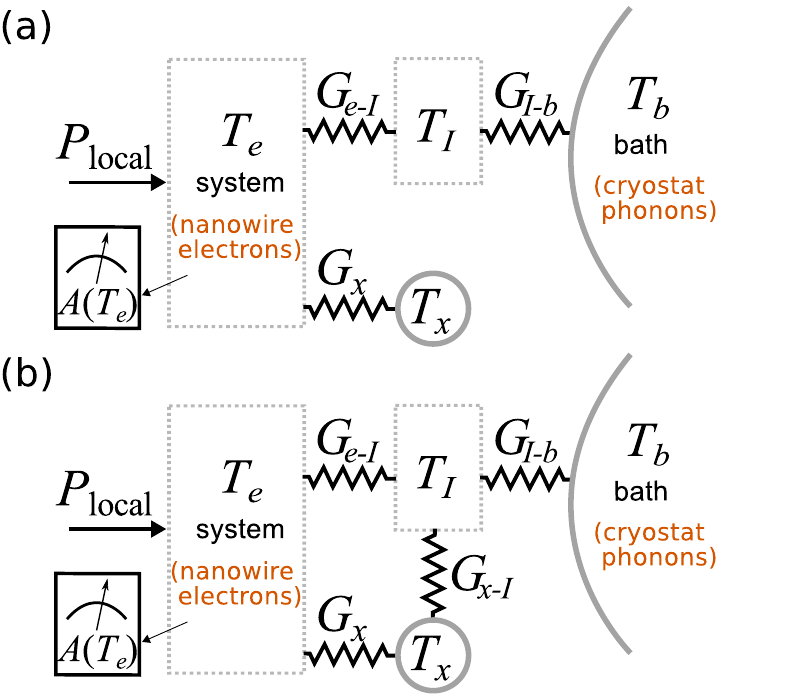}
\caption{(Color online)
  (a) Extended thermal model including an intermediate reservoir
  at temperature $T_I$ coupled to the nanowire electrons by $G_{e-I}$
  and to the cryostat phonons by $G_{I-b}$.
  (b) Same model as in (a) but with an additional parasitic heating power $T_x G_{x-I}$
  directly coupled to the intermediate reservoir.
  See the caption of Fig.~\ref{fig:setup}(a) for definitions of other symbols.
  \label{fig:caveats}}
\end{figure}

The main advantage of the isothermal technique is that it works
in the presence of an unknown parasitic heating power
without requiring calibration of the $T_e$ sensor.
However, the technique can lead to invalid conclusions
if the thermal model shown in Fig.~\ref{fig:setup}(a) does not accurately describe the system.
Here we present two more detailed thermal models (see Fig.~\ref{fig:caveats})
that illustrate why the isothermal technique can in certain circumstances lead to underestimation of the thermal energy fluctuations and the associated NEP.

We first note that the technique gives the total thermal conductance
of all the intermediate thermal links between the electron gas
and the phonons of the cryostat baseplate.
This leads to underestimation of $T_e$ fluctuations
if the bottleneck in heat conduction is $G_{I-b}$ between a large intermediate thermal reservoir
and the bath, rather than  $G_{e-I}$ between the electron gas and the intermediate reservoir
[see Fig.~\ref{fig:caveats}(a)].
As argued in the main text, the normal metal shadow of one of the leads
indeed constitutes such an intermediate reservoir for Sample \textsc{A}.
However, in general it is unlikely that $G_{I-b} < G_{e-I} \ll G_Q$
for macroscopic intermediate reservoirs.
On the other hand, mesoscopic bosonic reservoirs such as phonon\cite{pascal-prb-2013} or photon modes
tend to have negligible heat capacity compared to that of the electron gas.
In that case the simpler model in Fig.~\ref{fig:setup}(a) is appropriate
for assessing $T_e$ fluctuations, although $G_{e-b}$ is determined by $G_{I-b}$.

We also consider adding an additional parasitic heating power
to the intermediate reservoir [see Fig.~\ref{fig:caveats}(b)].
This model is more complex but potentially relevant at the
lowest bath temperatures where the constant parasitic power $T_x G_{x-I}$
may not be negligible compared to $T_b G_{I-b}$, even if $G_{I-b} \gg G_{e-I}$.
This scenario would be sufficient to prevent the intermediate temperature $T_I$ from following $T_b$,
leading to underestimation of the conductance between $T_e$ and $T_b$.
We do not have evidence that this phenomenon plays a noticeable role in our experiments,
but ultimately such speculation can be conclusively dismissed
only by direct measurement of the NEP.



\begin{figure}[t]
\includegraphics[width=8.6cm,keepaspectratio]{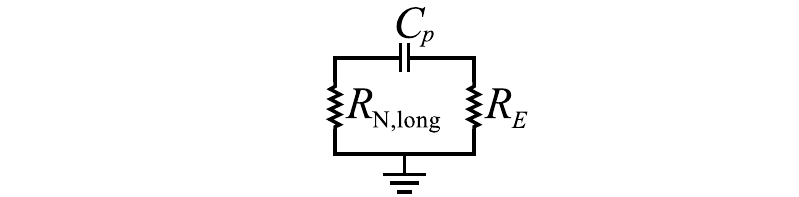}
\caption{
  Effective model for parasitic capacitive coupling $C_p$
  of the long junction ($R_{\textrm{N,long}}$)
  to a resistive environment ($R_E$).
  This model produces the upper (lower)
  black solid curve in Fig.~\ref{fig:g}
  with $R_{\textrm{N,long}}$, $R_E$, and $C_p$
  equal to $130 \, \Omega$, $120 \, \Omega$, and $300\mbox{ fF}$
  ($130 \, \Omega$, $1.5 \, \mbox{k}\Omega$, and $45\mbox{ fF}$),
  respectively.
  \label{fig:emmodel}}
\end{figure}

\section{Effective electromagnetic environment}\label{app:emenv}

Figure~\ref{fig:emmodel} shows a simple effective model
for the parasitic electron--photon coupling between the detector
and an unidentified electromagnetic environment.
We approximate the detector impedance as $R_{\textrm{N,long}}$,
although the real part of the short junction admittance
can contribute at high frequencies.
The capacitive coupling leads to better decoupling
of the sample from the environment as the thermal frequency $k_B T_b / h$ decreases,
leading to superlinear $T_b$ dependence as observed in Fig.~\ref{fig:g}.
We calculate the results numerically using formulas given by Pascal et al.\cite{pascal-prb-2011}

Although the values of the resistance of the environment $R_E$ and the parasitic capacitance $C_p$
have been chosen to fit the measured data, their magnitudes are reasonable (see Fig.~\ref{fig:emmodel}).
In dc mode, the first well defined high impedance ($1 \mbox{ k}\Omega$)
is centimeters away from the device on the printed circuit board,
making stray capacitances in the pF range realistic,
both between different signal lines
and the signal lines and the metallic sample enclosure.
In rf mode, $45 \mbox{ fF}$ is similar to the design value of the three-finger interdigitated
coupling capacitor of the on-chip CPW resonator (see G\"oppl et al.\cite{goppl-jap-2008})
used as a band-pass filter for the heating tone.
In both cases, the dissipation quantified by $R_E^{-1}$
may arise from dielectric losses in the intentional or unintentional capacitors
or from unintentionally generated shielding currents in normal metals.


\bibliography{uwavebolometer}

\begin{thebibliography}{70}%
\makeatletter
\providecommand \@ifxundefined [1]{%
 \@ifx{#1\undefined}
}%
\providecommand \@ifnum [1]{%
 \ifnum #1\expandafter \@firstoftwo
 \else \expandafter \@secondoftwo
 \fi
}%
\providecommand \@ifx [1]{%
 \ifx #1\expandafter \@firstoftwo
 \else \expandafter \@secondoftwo
 \fi
}%
\providecommand \natexlab [1]{#1}%
\providecommand \enquote  [1]{``#1''}%
\providecommand \bibnamefont  [1]{#1}%
\providecommand \bibfnamefont [1]{#1}%
\providecommand \citenamefont [1]{#1}%
\providecommand \href@noop [0]{\@secondoftwo}%
\providecommand \href [0]{\begingroup \@sanitize@url \@href}%
\providecommand \@href[1]{\@@startlink{#1}\@@href}%
\providecommand \@@href[1]{\endgroup#1\@@endlink}%
\providecommand \@sanitize@url [0]{\catcode `\\12\catcode `\$12\catcode
  `\&12\catcode `\#12\catcode `\^12\catcode `\_12\catcode `\%12\relax}%
\providecommand \@@startlink[1]{}%
\providecommand \@@endlink[0]{}%
\providecommand \url  [0]{\begingroup\@sanitize@url \@url }%
\providecommand \@url [1]{\endgroup\@href {#1}{\urlprefix }}%
\providecommand \urlprefix  [0]{URL }%
\providecommand \Eprint [0]{\href }%
\providecommand \doibase [0]{http://dx.doi.org/}%
\providecommand \selectlanguage [0]{\@gobble}%
\providecommand \bibinfo  [0]{\@secondoftwo}%
\providecommand \bibfield  [0]{\@secondoftwo}%
\providecommand \translation [1]{[#1]}%
\providecommand \BibitemOpen [0]{}%
\providecommand \bibitemStop [0]{}%
\providecommand \bibitemNoStop [0]{.\EOS\space}%
\providecommand \EOS [0]{\spacefactor3000\relax}%
\providecommand \BibitemShut  [1]{\csname bibitem#1\endcsname}%
\let\auto@bib@innerbib\@empty
\bibitem [{\citenamefont {Karasik}\ \emph {et~al.}(2011)\citenamefont
  {Karasik}, \citenamefont {Sergeev},\ and\ \citenamefont
  {Prober}}]{karasik-IEEETHZ-2011}%
  \BibitemOpen
  \bibfield  {author} {\bibinfo {author} {\bibfnamefont {B.}~\bibnamefont
  {Karasik}}, \bibinfo {author} {\bibfnamefont {A.}~\bibnamefont {Sergeev}}, \
  and\ \bibinfo {author} {\bibfnamefont {D.}~\bibnamefont {Prober}},\ }\href
  {\doibase 10.1109/TTHZ.2011.2159560} {\bibfield  {journal} {\bibinfo
  {journal} {IEEE Trans. Terahertz Sci.}\ }\textbf {\bibinfo {volume} {1}},\
  \bibinfo {pages} {97} (\bibinfo {year} {2011})}\BibitemShut {NoStop}%
\bibitem [{\citenamefont {Benford}\ and\ \citenamefont
  {Moseley}(2004)}]{benford-nimpr-2004}%
  \BibitemOpen
  \bibfield  {author} {\bibinfo {author} {\bibfnamefont {D.~J.}\ \bibnamefont
  {Benford}}\ and\ \bibinfo {author} {\bibfnamefont {S.~H.}\ \bibnamefont
  {Moseley}},\ }\href {\doibase 10.1016/j.nima.2003.11.295} {\bibfield
  {journal} {\bibinfo  {journal} {Nucl. Instrum. Methods Phys. Res., Sect. A}\
  }\textbf {\bibinfo {volume} {520}},\ \bibinfo {pages} {379} (\bibinfo {year}
  {2004})}\BibitemShut {NoStop}%
\bibitem [{\citenamefont {Blais}\ \emph {et~al.}(2004)\citenamefont {Blais},
  \citenamefont {Huang}, \citenamefont {Wallraff}, \citenamefont {Girvin},\
  and\ \citenamefont {Schoelkopf}}]{blais-pra-2004}%
  \BibitemOpen
  \bibfield  {author} {\bibinfo {author} {\bibfnamefont {A.}~\bibnamefont
  {Blais}}, \bibinfo {author} {\bibfnamefont {R.~S.}\ \bibnamefont {Huang}},
  \bibinfo {author} {\bibfnamefont {A.}~\bibnamefont {Wallraff}}, \bibinfo
  {author} {\bibfnamefont {S.~M.}\ \bibnamefont {Girvin}}, \ and\ \bibinfo
  {author} {\bibfnamefont {R.~J.}\ \bibnamefont {Schoelkopf}},\ }\href
  {\doibase 10.1103/physreva.69.062320} {\bibfield  {journal} {\bibinfo
  {journal} {Phys. Rev. A}\ }\textbf {\bibinfo {volume} {69}},\ \bibinfo
  {pages} {062320} (\bibinfo {year} {2004})}\BibitemShut {NoStop}%
\bibitem [{\citenamefont {Wallraff}\ \emph {et~al.}(2004)\citenamefont
  {Wallraff}, \citenamefont {Schuster}, \citenamefont {Blais}, \citenamefont
  {Frunzio}, \citenamefont {Huang}, \citenamefont {Majer}, \citenamefont
  {Kumar}, \citenamefont {Girvin},\ and\ \citenamefont
  {Schoelkopf}}]{wallraff-nature-2004}%
  \BibitemOpen
  \bibfield  {author} {\bibinfo {author} {\bibfnamefont {A.}~\bibnamefont
  {Wallraff}}, \bibinfo {author} {\bibfnamefont {D.~I.}\ \bibnamefont
  {Schuster}}, \bibinfo {author} {\bibfnamefont {A.}~\bibnamefont {Blais}},
  \bibinfo {author} {\bibfnamefont {L.}~\bibnamefont {Frunzio}}, \bibinfo
  {author} {\bibfnamefont {R.~S.}\ \bibnamefont {Huang}}, \bibinfo {author}
  {\bibfnamefont {J.}~\bibnamefont {Majer}}, \bibinfo {author} {\bibfnamefont
  {S.}~\bibnamefont {Kumar}}, \bibinfo {author} {\bibfnamefont {S.~M.}\
  \bibnamefont {Girvin}}, \ and\ \bibinfo {author} {\bibfnamefont {R.~J.}\
  \bibnamefont {Schoelkopf}},\ }\href {\doibase 10.1038/nature02851} {\bibfield
   {journal} {\bibinfo  {journal} {Nature (London)}\ }\textbf {\bibinfo
  {volume} {431}},\ \bibinfo {pages} {162} (\bibinfo {year}
  {2004})}\BibitemShut {NoStop}%
\bibitem [{\citenamefont {Riste}\ \emph {et~al.}(2013)\citenamefont {Riste},
  \citenamefont {Dukalski}, \citenamefont {Watson}, \citenamefont {de~Lange},
  \citenamefont {Tiggelman}, \citenamefont {Blanter}, \citenamefont {Lehnert},
  \citenamefont {Schouten},\ and\ \citenamefont {DiCarlo}}]{riste-nature-2013}%
  \BibitemOpen
  \bibfield  {author} {\bibinfo {author} {\bibfnamefont {D.}~\bibnamefont
  {Riste}}, \bibinfo {author} {\bibfnamefont {M.}~\bibnamefont {Dukalski}},
  \bibinfo {author} {\bibfnamefont {C.~A.}\ \bibnamefont {Watson}}, \bibinfo
  {author} {\bibfnamefont {G.}~\bibnamefont {de~Lange}}, \bibinfo {author}
  {\bibfnamefont {M.~J.}\ \bibnamefont {Tiggelman}}, \bibinfo {author}
  {\bibfnamefont {Y.}~\bibnamefont {Blanter}}, \bibinfo {author} {\bibfnamefont
  {K.~W.}\ \bibnamefont {Lehnert}}, \bibinfo {author} {\bibfnamefont {R.~N.}\
  \bibnamefont {Schouten}}, \ and\ \bibinfo {author} {\bibfnamefont
  {L.}~\bibnamefont {DiCarlo}},\ }\href {\doibase 10.1038/nature12513}
  {\bibfield  {journal} {\bibinfo  {journal} {Nature (London)}\ }\textbf
  {\bibinfo {volume} {502}},\ \bibinfo {pages} {350} (\bibinfo {year}
  {2013})}\BibitemShut {NoStop}%
\bibitem [{\citenamefont {Nation}\ \emph {et~al.}(2012)\citenamefont {Nation},
  \citenamefont {Johansson}, \citenamefont {Blencowe},\ and\ \citenamefont
  {Nori}}]{nation-rmp-2012}%
  \BibitemOpen
  \bibfield  {author} {\bibinfo {author} {\bibfnamefont {P.~D.}\ \bibnamefont
  {Nation}}, \bibinfo {author} {\bibfnamefont {J.~R.}\ \bibnamefont
  {Johansson}}, \bibinfo {author} {\bibfnamefont {M.~P.}\ \bibnamefont
  {Blencowe}}, \ and\ \bibinfo {author} {\bibfnamefont {F.}~\bibnamefont
  {Nori}},\ }\href {\doibase 10.1103/RevModPhys.84.1} {\bibfield  {journal}
  {\bibinfo  {journal} {Rev. Mod. Phys.}\ }\textbf {\bibinfo {volume} {84}},\
  \bibinfo {pages} {1} (\bibinfo {year} {2012})}\BibitemShut {NoStop}%
\bibitem [{\citenamefont {Hong}\ \emph {et~al.}(1987)\citenamefont {Hong},
  \citenamefont {Ou},\ and\ \citenamefont {Mandel}}]{hong-prl-1987}%
  \BibitemOpen
  \bibfield  {author} {\bibinfo {author} {\bibfnamefont {C.~K.}\ \bibnamefont
  {Hong}}, \bibinfo {author} {\bibfnamefont {Z.~Y.}\ \bibnamefont {Ou}}, \ and\
  \bibinfo {author} {\bibfnamefont {L.}~\bibnamefont {Mandel}},\ }\href
  {\doibase 10.1103/physrevlett.59.2044} {\bibfield  {journal} {\bibinfo
  {journal} {Phys. Rev. Lett.}\ }\textbf {\bibinfo {volume} {59}},\ \bibinfo
  {pages} {2044} (\bibinfo {year} {1987})}\BibitemShut {NoStop}%
\bibitem [{\citenamefont {Knill}\ \emph {et~al.}(2001)\citenamefont {Knill},
  \citenamefont {Laflamme},\ and\ \citenamefont {Milburn}}]{knill-nature-2001}%
  \BibitemOpen
  \bibfield  {author} {\bibinfo {author} {\bibfnamefont {E.}~\bibnamefont
  {Knill}}, \bibinfo {author} {\bibfnamefont {R.}~\bibnamefont {Laflamme}}, \
  and\ \bibinfo {author} {\bibfnamefont {G.~J.}\ \bibnamefont {Milburn}},\
  }\href {\doibase 10.1038/35051009} {\bibfield  {journal} {\bibinfo  {journal}
  {Nature (London)}\ }\textbf {\bibinfo {volume} {409}},\ \bibinfo {pages} {46}
  (\bibinfo {year} {2001})}\BibitemShut {NoStop}%
\bibitem [{\citenamefont {da~Silva}\ \emph {et~al.}(2010)\citenamefont
  {da~Silva}, \citenamefont {Bozyigit}, \citenamefont {Wallraff},\ and\
  \citenamefont {Blais}}]{dasilva-pra-2010}%
  \BibitemOpen
  \bibfield  {author} {\bibinfo {author} {\bibfnamefont {M.~P.}\ \bibnamefont
  {da~Silva}}, \bibinfo {author} {\bibfnamefont {D.}~\bibnamefont {Bozyigit}},
  \bibinfo {author} {\bibfnamefont {A.}~\bibnamefont {Wallraff}}, \ and\
  \bibinfo {author} {\bibfnamefont {A.}~\bibnamefont {Blais}},\ }\href
  {\doibase 10.1103/physreva.82.043804} {\bibfield  {journal} {\bibinfo
  {journal} {Phys. Rev. A}\ }\textbf {\bibinfo {volume} {82}},\ \bibinfo
  {pages} {043804} (\bibinfo {year} {2010})}\BibitemShut {NoStop}%
\bibitem [{\citenamefont {Lang}\ \emph {et~al.}(2013)\citenamefont {Lang},
  \citenamefont {Eichler}, \citenamefont {Steffen}, \citenamefont {Fink},
  \citenamefont {Woolley}, \citenamefont {Blais},\ and\ \citenamefont
  {Wallraff}}]{lang-nphys-2013}%
  \BibitemOpen
  \bibfield  {author} {\bibinfo {author} {\bibfnamefont {C.}~\bibnamefont
  {Lang}}, \bibinfo {author} {\bibfnamefont {C.}~\bibnamefont {Eichler}},
  \bibinfo {author} {\bibfnamefont {L.}~\bibnamefont {Steffen}}, \bibinfo
  {author} {\bibfnamefont {J.~M.}\ \bibnamefont {Fink}}, \bibinfo {author}
  {\bibfnamefont {M.~J.}\ \bibnamefont {Woolley}}, \bibinfo {author}
  {\bibfnamefont {A.}~\bibnamefont {Blais}}, \ and\ \bibinfo {author}
  {\bibfnamefont {A.}~\bibnamefont {Wallraff}},\ }\href {\doibase
  10.1038/nphys2612} {\bibfield  {journal} {\bibinfo  {journal} {Nature Phys.}\
  }\textbf {\bibinfo {volume} {9}},\ \bibinfo {pages} {345} (\bibinfo {year}
  {2013})}\BibitemShut {NoStop}%
\bibitem [{\citenamefont {Wei}\ \emph {et~al.}(2008)\citenamefont {Wei},
  \citenamefont {Olaya}, \citenamefont {Karasik}, \citenamefont {Pereverzev},
  \citenamefont {Sergeev},\ and\ \citenamefont
  {Gershenson}}]{wei-natnano-2008}%
  \BibitemOpen
  \bibfield  {author} {\bibinfo {author} {\bibfnamefont {J.}~\bibnamefont
  {Wei}}, \bibinfo {author} {\bibfnamefont {D.}~\bibnamefont {Olaya}}, \bibinfo
  {author} {\bibfnamefont {B.~S.}\ \bibnamefont {Karasik}}, \bibinfo {author}
  {\bibfnamefont {S.~V.}\ \bibnamefont {Pereverzev}}, \bibinfo {author}
  {\bibfnamefont {A.~V.}\ \bibnamefont {Sergeev}}, \ and\ \bibinfo {author}
  {\bibfnamefont {M.~E.}\ \bibnamefont {Gershenson}},\ }\href {\doibase
  10.1038/nnano.2008.173} {\bibfield  {journal} {\bibinfo  {journal} {Nature
  Nano.}\ }\textbf {\bibinfo {volume} {3}},\ \bibinfo {pages} {496} (\bibinfo
  {year} {2008})}\BibitemShut {NoStop}%
\bibitem [{\citenamefont {Karasik}\ and\ \citenamefont
  {Cantor}(2011)}]{karasik-apl-2011}%
  \BibitemOpen
  \bibfield  {author} {\bibinfo {author} {\bibfnamefont {B.~S.}\ \bibnamefont
  {Karasik}}\ and\ \bibinfo {author} {\bibfnamefont {R.}~\bibnamefont
  {Cantor}},\ }\href {\doibase 10.1063/1.3589367} {\bibfield  {journal}
  {\bibinfo  {journal} {Appl. Phys. Lett.}\ }\textbf {\bibinfo {volume} {98}},\
  \bibinfo {pages} {193503} (\bibinfo {year} {2011})}\BibitemShut {NoStop}%
\bibitem [{\citenamefont {Kenyon}\ \emph {et~al.}(2006)\citenamefont {Kenyon},
  \citenamefont {Day}, \citenamefont {Bradford}, \citenamefont {Bock},\ and\
  \citenamefont {Leduc}}]{kenyon-spie-2006}%
  \BibitemOpen
  \bibfield  {author} {\bibinfo {author} {\bibfnamefont {M.}~\bibnamefont
  {Kenyon}}, \bibinfo {author} {\bibfnamefont {P.~K.}\ \bibnamefont {Day}},
  \bibinfo {author} {\bibfnamefont {C.~M.}\ \bibnamefont {Bradford}}, \bibinfo
  {author} {\bibfnamefont {J.~J.}\ \bibnamefont {Bock}}, \ and\ \bibinfo
  {author} {\bibfnamefont {H.~G.}\ \bibnamefont {Leduc}},\ }\href {\doibase
  10.1117/12.672036} {\bibfield  {journal} {\bibinfo  {journal} {Proc. SPIE}\
  }\textbf {\bibinfo {volume} {6275}},\ \bibinfo {pages} {627508} (\bibinfo
  {year} {2006})}\BibitemShut {NoStop}%
\bibitem [{\citenamefont {Santavicca}\ \emph {et~al.}(2010)\citenamefont
  {Santavicca}, \citenamefont {Reulet}, \citenamefont {Karasik}, \citenamefont
  {Pereverzev}, \citenamefont {Olaya}, \citenamefont {Gershenson},
  \citenamefont {Frunzio},\ and\ \citenamefont {Prober}}]{santavicca-apl-2010}%
  \BibitemOpen
  \bibfield  {author} {\bibinfo {author} {\bibfnamefont {D.~F.}\ \bibnamefont
  {Santavicca}}, \bibinfo {author} {\bibfnamefont {B.}~\bibnamefont {Reulet}},
  \bibinfo {author} {\bibfnamefont {B.~S.}\ \bibnamefont {Karasik}}, \bibinfo
  {author} {\bibfnamefont {S.~V.}\ \bibnamefont {Pereverzev}}, \bibinfo
  {author} {\bibfnamefont {D.}~\bibnamefont {Olaya}}, \bibinfo {author}
  {\bibfnamefont {M.~E.}\ \bibnamefont {Gershenson}}, \bibinfo {author}
  {\bibfnamefont {L.}~\bibnamefont {Frunzio}}, \ and\ \bibinfo {author}
  {\bibfnamefont {D.~E.}\ \bibnamefont {Prober}},\ }\href {\doibase
  10.1063/1.3336008} {\bibfield  {journal} {\bibinfo  {journal} {Appl. Phys.
  Lett.}\ }\textbf {\bibinfo {volume} {96}},\ \bibinfo {pages} {083505}
  (\bibinfo {year} {2010})}\BibitemShut {NoStop}%
\bibitem [{\citenamefont {Karasik}\ \emph {et~al.}(2012)\citenamefont
  {Karasik}, \citenamefont {Pereverzev}, \citenamefont {Soibel}, \citenamefont
  {Santavicca}, \citenamefont {Prober}, \citenamefont {Olaya},\ and\
  \citenamefont {Gershenson}}]{karasik-apl-2012}%
  \BibitemOpen
  \bibfield  {author} {\bibinfo {author} {\bibfnamefont {B.~S.}\ \bibnamefont
  {Karasik}}, \bibinfo {author} {\bibfnamefont {S.~V.}\ \bibnamefont
  {Pereverzev}}, \bibinfo {author} {\bibfnamefont {A.}~\bibnamefont {Soibel}},
  \bibinfo {author} {\bibfnamefont {D.~F.}\ \bibnamefont {Santavicca}},
  \bibinfo {author} {\bibfnamefont {D.~E.}\ \bibnamefont {Prober}}, \bibinfo
  {author} {\bibfnamefont {D.}~\bibnamefont {Olaya}}, \ and\ \bibinfo {author}
  {\bibfnamefont {M.~E.}\ \bibnamefont {Gershenson}},\ }\href {\doibase
  10.1063/1.4739839} {\bibfield  {journal} {\bibinfo  {journal} {Appl. Phys.
  Lett.}\ }\textbf {\bibinfo {volume} {101}},\ \bibinfo {pages} {052601}
  (\bibinfo {year} {2012})}\BibitemShut {NoStop}%
\bibitem [{\citenamefont {Day}\ \emph {et~al.}(2003)\citenamefont {Day},
  \citenamefont {LeDuc}, \citenamefont {Mazin}, \citenamefont {Vayonakis},\
  and\ \citenamefont {Zmuidzinas}}]{day-nature-2003}%
  \BibitemOpen
  \bibfield  {author} {\bibinfo {author} {\bibfnamefont {P.~K.}\ \bibnamefont
  {Day}}, \bibinfo {author} {\bibfnamefont {H.~G.}\ \bibnamefont {LeDuc}},
  \bibinfo {author} {\bibfnamefont {B.~A.}\ \bibnamefont {Mazin}}, \bibinfo
  {author} {\bibfnamefont {A.}~\bibnamefont {Vayonakis}}, \ and\ \bibinfo
  {author} {\bibfnamefont {J.}~\bibnamefont {Zmuidzinas}},\ }\href {\doibase
  10.1038/nature02037} {\bibfield  {journal} {\bibinfo  {journal} {Nature
  (London)}\ }\textbf {\bibinfo {volume} {425}},\ \bibinfo {pages} {817}
  (\bibinfo {year} {2003})}\BibitemShut {NoStop}%
\bibitem [{\citenamefont {Janssen}\ \emph {et~al.}(2013)\citenamefont
  {Janssen}, \citenamefont {Baselmans}, \citenamefont {Endo}, \citenamefont
  {Ferrari}, \citenamefont {Yates}, \citenamefont {Baryshev},\ and\
  \citenamefont {Klapwijk}}]{janssen-apl-2013}%
  \BibitemOpen
  \bibfield  {author} {\bibinfo {author} {\bibfnamefont {R.~M.~J.}\
  \bibnamefont {Janssen}}, \bibinfo {author} {\bibfnamefont {J.~J.~A.}\
  \bibnamefont {Baselmans}}, \bibinfo {author} {\bibfnamefont {A.}~\bibnamefont
  {Endo}}, \bibinfo {author} {\bibfnamefont {L.}~\bibnamefont {Ferrari}},
  \bibinfo {author} {\bibfnamefont {S.~J.~C.}\ \bibnamefont {Yates}}, \bibinfo
  {author} {\bibfnamefont {A.~M.}\ \bibnamefont {Baryshev}}, \ and\ \bibinfo
  {author} {\bibfnamefont {T.~M.}\ \bibnamefont {Klapwijk}},\ }\href {\doibase
  10.1063/1.4829657} {\bibfield  {journal} {\bibinfo  {journal} {Appl. Phys.
  Lett.}\ }\textbf {\bibinfo {volume} {103}},\ \bibinfo {pages} {203503}
  (\bibinfo {year} {2013})}\BibitemShut {NoStop}%
\bibitem [{\citenamefont {de~Visser}\ \emph {et~al.}(2014)\citenamefont
  {de~Visser}, \citenamefont {Baselmans}, \citenamefont {Bueno}, \citenamefont
  {Llombart},\ and\ \citenamefont {Klapwijk}}]{devisser-natcom-2014}%
  \BibitemOpen
  \bibfield  {author} {\bibinfo {author} {\bibfnamefont {P.~J.}\ \bibnamefont
  {de~Visser}}, \bibinfo {author} {\bibfnamefont {J.~J.~A.}\ \bibnamefont
  {Baselmans}}, \bibinfo {author} {\bibfnamefont {J.}~\bibnamefont {Bueno}},
  \bibinfo {author} {\bibfnamefont {N.}~\bibnamefont {Llombart}}, \ and\
  \bibinfo {author} {\bibfnamefont {T.~M.}\ \bibnamefont {Klapwijk}},\ }\href
  {\doibase 10.1038/ncomms4130} {\bibfield  {journal} {\bibinfo  {journal}
  {Nature Commun.}\ }\textbf {\bibinfo {volume} {5}},\ \bibinfo {pages} {3130}
  (\bibinfo {year} {2014})}\BibitemShut {NoStop}%
\bibitem [{\citenamefont {Nahum}\ and\ \citenamefont
  {Martinis}(1994)}]{nahum-physicab-1994}%
  \BibitemOpen
  \bibfield  {author} {\bibinfo {author} {\bibfnamefont {M.}~\bibnamefont
  {Nahum}}\ and\ \bibinfo {author} {\bibfnamefont {J.}~\bibnamefont
  {Martinis}},\ }\href {\doibase 10.1016/0921-4526(94)90384-0} {\bibfield
  {journal} {\bibinfo  {journal} {Physica B}\ }\textbf {\bibinfo {volume}
  {194}},\ \bibinfo {pages} {109} (\bibinfo {year} {1994})}\BibitemShut
  {NoStop}%
\bibitem [{\citenamefont {Schmidt}\ \emph {et~al.}(2003)\citenamefont
  {Schmidt}, \citenamefont {Yung},\ and\ \citenamefont
  {Cleland}}]{schmidt-apl-2003}%
  \BibitemOpen
  \bibfield  {author} {\bibinfo {author} {\bibfnamefont {D.~R.}\ \bibnamefont
  {Schmidt}}, \bibinfo {author} {\bibfnamefont {C.~S.}\ \bibnamefont {Yung}}, \
  and\ \bibinfo {author} {\bibfnamefont {A.~N.}\ \bibnamefont {Cleland}},\
  }\href {\doibase 10.1063/1.1597983} {\bibfield  {journal} {\bibinfo
  {journal} {Appl. Phys. Lett.}\ }\textbf {\bibinfo {volume} {83}},\ \bibinfo
  {pages} {1002} (\bibinfo {year} {2003})}\BibitemShut {NoStop}%
\bibitem [{\citenamefont {Schmidt}\ \emph {et~al.}(2005)\citenamefont
  {Schmidt}, \citenamefont {Lehnert}, \citenamefont {Clark}, \citenamefont
  {Duncan}, \citenamefont {Irwin}, \citenamefont {Miller},\ and\ \citenamefont
  {Ullom}}]{schmidt-apl-2005}%
  \BibitemOpen
  \bibfield  {author} {\bibinfo {author} {\bibfnamefont {D.~R.}\ \bibnamefont
  {Schmidt}}, \bibinfo {author} {\bibfnamefont {K.~W.}\ \bibnamefont
  {Lehnert}}, \bibinfo {author} {\bibfnamefont {A.~M.}\ \bibnamefont {Clark}},
  \bibinfo {author} {\bibfnamefont {W.~D.}\ \bibnamefont {Duncan}}, \bibinfo
  {author} {\bibfnamefont {K.~D.}\ \bibnamefont {Irwin}}, \bibinfo {author}
  {\bibfnamefont {N.}~\bibnamefont {Miller}}, \ and\ \bibinfo {author}
  {\bibfnamefont {J.~N.}\ \bibnamefont {Ullom}},\ }\href {\doibase
  10.1063/1.1855411} {\bibfield  {journal} {\bibinfo  {journal} {Appl. Phys.
  Lett.}\ }\textbf {\bibinfo {volume} {86}},\ \bibinfo {pages} {053505}
  (\bibinfo {year} {2005})}\BibitemShut {NoStop}%
\bibitem [{\citenamefont {Prober}\ \emph {et~al.}(2007)\citenamefont {Prober},
  \citenamefont {Teufel}, \citenamefont {Wilson}, \citenamefont {Frunzio},
  \citenamefont {Shen}, \citenamefont {Schoelkopf}, \citenamefont {Stevenson},\
  and\ \citenamefont {Wollack}}]{prober-ieee-2007}%
  \BibitemOpen
  \bibfield  {author} {\bibinfo {author} {\bibfnamefont {D.~E.}\ \bibnamefont
  {Prober}}, \bibinfo {author} {\bibfnamefont {J.~D.}\ \bibnamefont {Teufel}},
  \bibinfo {author} {\bibfnamefont {C.~M.}\ \bibnamefont {Wilson}}, \bibinfo
  {author} {\bibfnamefont {L.}~\bibnamefont {Frunzio}}, \bibinfo {author}
  {\bibfnamefont {M.}~\bibnamefont {Shen}}, \bibinfo {author} {\bibfnamefont
  {R.~J.}\ \bibnamefont {Schoelkopf}}, \bibinfo {author} {\bibfnamefont
  {T.~R.}\ \bibnamefont {Stevenson}}, \ and\ \bibinfo {author} {\bibfnamefont
  {E.~J.}\ \bibnamefont {Wollack}},\ }\href {\doibase 10.1109/tasc.2007.897397}
  {\bibfield  {journal} {\bibinfo  {journal} {IEEE Trans. Appl. Supercond.}\
  }\textbf {\bibinfo {volume} {17}},\ \bibinfo {pages} {241} (\bibinfo {year}
  {2007})}\BibitemShut {NoStop}%
\bibitem [{\citenamefont {Komiyama}\ \emph {et~al.}(2000)\citenamefont
  {Komiyama}, \citenamefont {Astafiev}, \citenamefont {Antonov}, \citenamefont
  {Kutsuwa},\ and\ \citenamefont {Hirai}}]{komiyama-nature-2000}%
  \BibitemOpen
  \bibfield  {author} {\bibinfo {author} {\bibfnamefont {S.}~\bibnamefont
  {Komiyama}}, \bibinfo {author} {\bibfnamefont {O.}~\bibnamefont {Astafiev}},
  \bibinfo {author} {\bibfnamefont {V.}~\bibnamefont {Antonov}}, \bibinfo
  {author} {\bibfnamefont {T.}~\bibnamefont {Kutsuwa}}, \ and\ \bibinfo
  {author} {\bibfnamefont {H.}~\bibnamefont {Hirai}},\ }\href {\doibase
  10.1038/35000166} {\bibfield  {journal} {\bibinfo  {journal} {Nature
  (London)}\ }\textbf {\bibinfo {volume} {403}},\ \bibinfo {pages} {405}
  (\bibinfo {year} {2000})}\BibitemShut {NoStop}%
\bibitem [{\citenamefont {Hashiba}\ \emph {et~al.}(2010)\citenamefont
  {Hashiba}, \citenamefont {Antonov}, \citenamefont {Kulik}, \citenamefont
  {Tzalenchuk},\ and\ \citenamefont {Komiyama}}]{hashiba-nanotech-2010}%
  \BibitemOpen
  \bibfield  {author} {\bibinfo {author} {\bibfnamefont {H.}~\bibnamefont
  {Hashiba}}, \bibinfo {author} {\bibfnamefont {V.}~\bibnamefont {Antonov}},
  \bibinfo {author} {\bibfnamefont {L.}~\bibnamefont {Kulik}}, \bibinfo
  {author} {\bibfnamefont {A.}~\bibnamefont {Tzalenchuk}}, \ and\ \bibinfo
  {author} {\bibfnamefont {S.}~\bibnamefont {Komiyama}},\ }\href {\doibase
  10.1088/0957-4484/21/16/165203} {\bibfield  {journal} {\bibinfo  {journal}
  {Nanotechnol.}\ }\textbf {\bibinfo {volume} {21}},\ \bibinfo {pages} {165203}
  (\bibinfo {year} {2010})}\BibitemShut {NoStop}%
\bibitem [{\citenamefont {Komiyama}(2011)}]{komiyama-ieee-2011}%
  \BibitemOpen
  \bibfield  {author} {\bibinfo {author} {\bibfnamefont {S.}~\bibnamefont
  {Komiyama}},\ }\href {\doibase 10.1109/jstqe.2010.2048893} {\bibfield
  {journal} {\bibinfo  {journal} {IEEE J. Sel. Top. Quant. Electr.}\ }\textbf
  {\bibinfo {volume} {17}},\ \bibinfo {pages} {54} (\bibinfo {year}
  {2011})}\BibitemShut {NoStop}%
\bibitem [{\citenamefont {Fong}\ and\ \citenamefont
  {Schwab}(2012)}]{fong-prx-2012}%
  \BibitemOpen
  \bibfield  {author} {\bibinfo {author} {\bibfnamefont {K.~C.}\ \bibnamefont
  {Fong}}\ and\ \bibinfo {author} {\bibfnamefont {K.~C.}\ \bibnamefont
  {Schwab}},\ }\href {\doibase 10.1103/PhysRevX.2.031006} {\bibfield  {journal}
  {\bibinfo  {journal} {Phys. Rev. X}\ }\textbf {\bibinfo {volume} {2}},\
  \bibinfo {pages} {031006} (\bibinfo {year} {2012})}\BibitemShut {NoStop}%
\bibitem [{\citenamefont {Yan}\ \emph {et~al.}(2012)\citenamefont {Yan},
  \citenamefont {Kim}, \citenamefont {Elle}, \citenamefont {Sushkov},
  \citenamefont {Jenkins}, \citenamefont {Milchberg}, \citenamefont {Fuhrer},\
  and\ \citenamefont {Drew}}]{yan-natnano-2012}%
  \BibitemOpen
  \bibfield  {author} {\bibinfo {author} {\bibfnamefont {J.}~\bibnamefont
  {Yan}}, \bibinfo {author} {\bibfnamefont {M.-H.}\ \bibnamefont {Kim}},
  \bibinfo {author} {\bibfnamefont {J.~A.}\ \bibnamefont {Elle}}, \bibinfo
  {author} {\bibfnamefont {A.~B.}\ \bibnamefont {Sushkov}}, \bibinfo {author}
  {\bibfnamefont {G.~S.}\ \bibnamefont {Jenkins}}, \bibinfo {author}
  {\bibfnamefont {H.~M.}\ \bibnamefont {Milchberg}}, \bibinfo {author}
  {\bibfnamefont {M.~S.}\ \bibnamefont {Fuhrer}}, \ and\ \bibinfo {author}
  {\bibfnamefont {H.~D.}\ \bibnamefont {Drew}},\ }\href {\doibase
  10.1038/nnano.2012.88} {\bibfield  {journal} {\bibinfo  {journal} {Nature
  Nano.}\ }\textbf {\bibinfo {volume} {7}},\ \bibinfo {pages} {472} (\bibinfo
  {year} {2012})}\BibitemShut {NoStop}%
\bibitem [{\citenamefont {Mittendorff}\ \emph {et~al.}(2013)\citenamefont
  {Mittendorff}, \citenamefont {Winnerl}, \citenamefont {Kamann}, \citenamefont
  {Eroms}, \citenamefont {Weiss}, \citenamefont {Schneider},\ and\
  \citenamefont {Helm}}]{mittendorff-apl-2013}%
  \BibitemOpen
  \bibfield  {author} {\bibinfo {author} {\bibfnamefont {M.}~\bibnamefont
  {Mittendorff}}, \bibinfo {author} {\bibfnamefont {S.}~\bibnamefont
  {Winnerl}}, \bibinfo {author} {\bibfnamefont {J.}~\bibnamefont {Kamann}},
  \bibinfo {author} {\bibfnamefont {J.}~\bibnamefont {Eroms}}, \bibinfo
  {author} {\bibfnamefont {D.}~\bibnamefont {Weiss}}, \bibinfo {author}
  {\bibfnamefont {H.}~\bibnamefont {Schneider}}, \ and\ \bibinfo {author}
  {\bibfnamefont {M.}~\bibnamefont {Helm}},\ }\href {\doibase
  10.1063/1.4813621} {\bibfield  {journal} {\bibinfo  {journal} {Appl. Phys.
  Lett.}\ }\textbf {\bibinfo {volume} {103}},\ \bibinfo {pages} {021113}
  (\bibinfo {year} {2013})}\BibitemShut {NoStop}%
\bibitem [{\citenamefont {McKitterick}\ \emph {et~al.}(2013)\citenamefont
  {McKitterick}, \citenamefont {Prober},\ and\ \citenamefont
  {Karasik}}]{mckitterick-jap-2013}%
  \BibitemOpen
  \bibfield  {author} {\bibinfo {author} {\bibfnamefont {C.~B.}\ \bibnamefont
  {McKitterick}}, \bibinfo {author} {\bibfnamefont {D.~E.}\ \bibnamefont
  {Prober}}, \ and\ \bibinfo {author} {\bibfnamefont {B.~S.}\ \bibnamefont
  {Karasik}},\ }\href {\doibase 10.1063/1.4789360} {\bibfield  {journal}
  {\bibinfo  {journal} {J. Appl. Phys.}\ }\textbf {\bibinfo {volume} {113}},\
  \bibinfo {eid} {044512} (\bibinfo {year} {2013})}\BibitemShut {NoStop}%
\bibitem [{\citenamefont {Kawano}\ \emph {et~al.}(2009)\citenamefont {Kawano},
  \citenamefont {Uchida},\ and\ \citenamefont {Ishibashi}}]{kawano-apl-2009}%
  \BibitemOpen
  \bibfield  {author} {\bibinfo {author} {\bibfnamefont {Y.}~\bibnamefont
  {Kawano}}, \bibinfo {author} {\bibfnamefont {T.}~\bibnamefont {Uchida}}, \
  and\ \bibinfo {author} {\bibfnamefont {K.}~\bibnamefont {Ishibashi}},\ }\href
  {\doibase 10.1063/1.3205125} {\bibfield  {journal} {\bibinfo  {journal}
  {Appl. Phys. Lett.}\ }\textbf {\bibinfo {volume} {95}},\ \bibinfo {pages}
  {083123} (\bibinfo {year} {2009})}\BibitemShut {NoStop}%
\bibitem [{\citenamefont {Richards}\ \emph {et~al.}(1980)\citenamefont
  {Richards}, \citenamefont {Shen}, \citenamefont {Harris},\ and\ \citenamefont
  {Lloyd}}]{richards-apl-1980}%
  \BibitemOpen
  \bibfield  {author} {\bibinfo {author} {\bibfnamefont {P.~L.}\ \bibnamefont
  {Richards}}, \bibinfo {author} {\bibfnamefont {T.~M.}\ \bibnamefont {Shen}},
  \bibinfo {author} {\bibfnamefont {R.~E.}\ \bibnamefont {Harris}}, \ and\
  \bibinfo {author} {\bibfnamefont {F.~L.}\ \bibnamefont {Lloyd}},\ }\href
  {\doibase 10.1063/1.91514} {\bibfield  {journal} {\bibinfo  {journal} {Appl.
  Phys. Lett.}\ }\textbf {\bibinfo {volume} {36}},\ \bibinfo {pages} {480}
  (\bibinfo {year} {1980})}\BibitemShut {NoStop}%
\bibitem [{\citenamefont {Bueno}\ \emph {et~al.}(2010)\citenamefont {Bueno},
  \citenamefont {Shaw}, \citenamefont {Day},\ and\ \citenamefont
  {Echternach}}]{bueno-apl-2010}%
  \BibitemOpen
  \bibfield  {author} {\bibinfo {author} {\bibfnamefont {J.}~\bibnamefont
  {Bueno}}, \bibinfo {author} {\bibfnamefont {M.~D.}\ \bibnamefont {Shaw}},
  \bibinfo {author} {\bibfnamefont {P.~K.}\ \bibnamefont {Day}}, \ and\
  \bibinfo {author} {\bibfnamefont {P.~M.}\ \bibnamefont {Echternach}},\ }\href
  {\doibase 10.1063/1.3339163} {\bibfield  {journal} {\bibinfo  {journal}
  {Appl. Phys. Lett.}\ }\textbf {\bibinfo {volume} {96}},\ \bibinfo {pages}
  {103503} (\bibinfo {year} {2010})}\BibitemShut {NoStop}%
\bibitem [{\citenamefont {Echternach}\ \emph {et~al.}(2013)\citenamefont
  {Echternach}, \citenamefont {Stone}, \citenamefont {Bradford}, \citenamefont
  {Day}, \citenamefont {Wilson}, \citenamefont {Megerian}, \citenamefont
  {Llombart},\ and\ \citenamefont {Bueno}}]{echternach-apl-2013}%
  \BibitemOpen
  \bibfield  {author} {\bibinfo {author} {\bibfnamefont {P.~M.}\ \bibnamefont
  {Echternach}}, \bibinfo {author} {\bibfnamefont {K.~J.}\ \bibnamefont
  {Stone}}, \bibinfo {author} {\bibfnamefont {C.~M.}\ \bibnamefont {Bradford}},
  \bibinfo {author} {\bibfnamefont {P.~K.}\ \bibnamefont {Day}}, \bibinfo
  {author} {\bibfnamefont {D.~W.}\ \bibnamefont {Wilson}}, \bibinfo {author}
  {\bibfnamefont {K.~G.}\ \bibnamefont {Megerian}}, \bibinfo {author}
  {\bibfnamefont {N.}~\bibnamefont {Llombart}}, \ and\ \bibinfo {author}
  {\bibfnamefont {J.}~\bibnamefont {Bueno}},\ }\href {\doibase
  10.1063/1.4817585} {\bibfield  {journal} {\bibinfo  {journal} {Appl. Phys.
  Lett.}\ }\textbf {\bibinfo {volume} {103}},\ \bibinfo {pages} {053510}
  (\bibinfo {year} {2013})}\BibitemShut {NoStop}%
\bibitem [{\citenamefont {Sadleir}\ \emph {et~al.}(2010)\citenamefont
  {Sadleir}, \citenamefont {Smith}, \citenamefont {Bandler}, \citenamefont
  {Chervenak},\ and\ \citenamefont {Clem}}]{sadleir-prl-2010}%
  \BibitemOpen
  \bibfield  {author} {\bibinfo {author} {\bibfnamefont {J.~E.}\ \bibnamefont
  {Sadleir}}, \bibinfo {author} {\bibfnamefont {S.~J.}\ \bibnamefont {Smith}},
  \bibinfo {author} {\bibfnamefont {S.~R.}\ \bibnamefont {Bandler}}, \bibinfo
  {author} {\bibfnamefont {J.~A.}\ \bibnamefont {Chervenak}}, \ and\ \bibinfo
  {author} {\bibfnamefont {J.~R.}\ \bibnamefont {Clem}},\ }\href {\doibase
  10.1103/physrevlett.104.047003} {\bibfield  {journal} {\bibinfo  {journal}
  {Phys. Rev. Lett.}\ }\textbf {\bibinfo {volume} {104}},\ \bibinfo {pages}
  {047003} (\bibinfo {year} {2010})}\BibitemShut {NoStop}%
\bibitem [{\citenamefont {Wellstood}\ \emph {et~al.}(1994)\citenamefont
  {Wellstood}, \citenamefont {Urbina},\ and\ \citenamefont
  {Clarke}}]{wellstood-prb-1994}%
  \BibitemOpen
  \bibfield  {author} {\bibinfo {author} {\bibfnamefont {F.~C.}\ \bibnamefont
  {Wellstood}}, \bibinfo {author} {\bibfnamefont {C.}~\bibnamefont {Urbina}}, \
  and\ \bibinfo {author} {\bibfnamefont {J.}~\bibnamefont {Clarke}},\ }\href
  {\doibase 10.1103/PhysRevB.49.5942} {\bibfield  {journal} {\bibinfo
  {journal} {Phys. Rev. B}\ }\textbf {\bibinfo {volume} {49}},\ \bibinfo
  {pages} {5942} (\bibinfo {year} {1994})}\BibitemShut {NoStop}%
\bibitem [{\citenamefont {Giazotto}\ \emph {et~al.}(2008)\citenamefont
  {Giazotto}, \citenamefont {Heikkil\"{a}}, \citenamefont {Pepe}, \citenamefont
  {Helist\"{o}}, \citenamefont {Luukanen},\ and\ \citenamefont
  {Pekola}}]{giazotto-apl-2008}%
  \BibitemOpen
  \bibfield  {author} {\bibinfo {author} {\bibfnamefont {F.}~\bibnamefont
  {Giazotto}}, \bibinfo {author} {\bibfnamefont {T.~T.}\ \bibnamefont
  {Heikkil\"{a}}}, \bibinfo {author} {\bibfnamefont {G.~P.}\ \bibnamefont
  {Pepe}}, \bibinfo {author} {\bibfnamefont {P.}~\bibnamefont {Helist\"{o}}},
  \bibinfo {author} {\bibfnamefont {A.}~\bibnamefont {Luukanen}}, \ and\
  \bibinfo {author} {\bibfnamefont {J.~P.}\ \bibnamefont {Pekola}},\ }\href
  {\doibase http://dx.doi.org/10.1063/1.2908922} {\bibfield  {journal}
  {\bibinfo  {journal} {Appl. Phys. Lett.}\ }\textbf {\bibinfo {volume} {92}},\
  \bibinfo {pages} {162507} (\bibinfo {year} {2008})}\BibitemShut {NoStop}%
\bibitem [{\citenamefont {Voutilainen}\ \emph {et~al.}(2010)\citenamefont
  {Voutilainen}, \citenamefont {Laakso},\ and\ \citenamefont
  {Heikkil\"{a}}}]{voutilainen-jap-2010}%
  \BibitemOpen
  \bibfield  {author} {\bibinfo {author} {\bibfnamefont {J.}~\bibnamefont
  {Voutilainen}}, \bibinfo {author} {\bibfnamefont {M.~A.}\ \bibnamefont
  {Laakso}}, \ and\ \bibinfo {author} {\bibfnamefont {T.~T.}\ \bibnamefont
  {Heikkil\"{a}}},\ }\href {\doibase 10.1063/1.3354042} {\bibfield  {journal}
  {\bibinfo  {journal} {J. Appl. Phys.}\ }\textbf {\bibinfo {volume} {107}},\
  \bibinfo {pages} {064508} (\bibinfo {year} {2010})}\BibitemShut {NoStop}%
\bibitem [{\citenamefont {Mather}(1982)}]{mather-ao-1982}%
  \BibitemOpen
  \bibfield  {author} {\bibinfo {author} {\bibfnamefont {J.~C.}\ \bibnamefont
  {Mather}},\ }\href {\doibase 10.1364/AO.21.001125} {\bibfield  {journal}
  {\bibinfo  {journal} {Appl. Opt.}\ }\textbf {\bibinfo {volume} {21}},\
  \bibinfo {pages} {1125} (\bibinfo {year} {1982})}\BibitemShut {NoStop}%
\bibitem [{\citenamefont {Moseley}\ \emph {et~al.}(1984)\citenamefont
  {Moseley}, \citenamefont {Mather},\ and\ \citenamefont
  {McCammon}}]{moseley-jap-1984}%
  \BibitemOpen
  \bibfield  {author} {\bibinfo {author} {\bibfnamefont {S.~H.}\ \bibnamefont
  {Moseley}}, \bibinfo {author} {\bibfnamefont {J.~C.}\ \bibnamefont {Mather}},
  \ and\ \bibinfo {author} {\bibfnamefont {D.}~\bibnamefont {McCammon}},\
  }\href {\doibase 10.1063/1.334129} {\bibfield  {journal} {\bibinfo  {journal}
  {J. Appl. Phys.}\ }\textbf {\bibinfo {volume} {56}},\ \bibinfo {pages} {1257}
  (\bibinfo {year} {1984})}\BibitemShut {NoStop}%
\bibitem [{\citenamefont {Booth}\ \emph {et~al.}(1996)\citenamefont {Booth},
  \citenamefont {Cabrera},\ and\ \citenamefont {Fiorini}}]{booth-arnps-1996}%
  \BibitemOpen
  \bibfield  {author} {\bibinfo {author} {\bibfnamefont {N.~E.}\ \bibnamefont
  {Booth}}, \bibinfo {author} {\bibfnamefont {B.}~\bibnamefont {Cabrera}}, \
  and\ \bibinfo {author} {\bibfnamefont {E.}~\bibnamefont {Fiorini}},\ }\href
  {\doibase 10.1146/annurev.nucl.46.1.471} {\bibfield  {journal} {\bibinfo
  {journal} {Annu. Rev. Nucl. Part. Sci.}\ }\textbf {\bibinfo {volume} {46}},\
  \bibinfo {pages} {471} (\bibinfo {year} {1996})}\BibitemShut {NoStop}%
\bibitem [{\citenamefont {Pendry}(1983)}]{pendry-1983}%
  \BibitemOpen
  \bibfield  {author} {\bibinfo {author} {\bibfnamefont {J.~B.}\ \bibnamefont
  {Pendry}},\ }\href {\doibase 10.1088/0305-4470/16/10/012} {\bibfield
  {journal} {\bibinfo  {journal} {J. Phys. A}\ }\textbf {\bibinfo {volume}
  {16}},\ \bibinfo {pages} {2161} (\bibinfo {year} {1983})}\BibitemShut
  {NoStop}%
\bibitem [{\citenamefont {Schwab}\ \emph {et~al.}(2000)\citenamefont {Schwab},
  \citenamefont {Henriksen}, \citenamefont {Worlock},\ and\ \citenamefont
  {Roukes}}]{schwab-nature-2000}%
  \BibitemOpen
  \bibfield  {author} {\bibinfo {author} {\bibfnamefont {K.}~\bibnamefont
  {Schwab}}, \bibinfo {author} {\bibfnamefont {E.~A.}\ \bibnamefont
  {Henriksen}}, \bibinfo {author} {\bibfnamefont {J.~M.}\ \bibnamefont
  {Worlock}}, \ and\ \bibinfo {author} {\bibfnamefont {M.~L.}\ \bibnamefont
  {Roukes}},\ }\href {\doibase 10.1038/35010065} {\bibfield  {journal}
  {\bibinfo  {journal} {Nature (London)}\ }\textbf {\bibinfo {volume} {404}},\
  \bibinfo {pages} {974} (\bibinfo {year} {2000})}\BibitemShut {NoStop}%
\bibitem [{\citenamefont {Schmidt}\ \emph {et~al.}(2004)\citenamefont
  {Schmidt}, \citenamefont {Schoelkopf},\ and\ \citenamefont
  {Cleland}}]{schmidt-prl-2004}%
  \BibitemOpen
  \bibfield  {author} {\bibinfo {author} {\bibfnamefont {D.~R.}\ \bibnamefont
  {Schmidt}}, \bibinfo {author} {\bibfnamefont {R.~J.}\ \bibnamefont
  {Schoelkopf}}, \ and\ \bibinfo {author} {\bibfnamefont {A.~N.}\ \bibnamefont
  {Cleland}},\ }\href {\doibase 10.1103/PhysRevLett.93.045901} {\bibfield
  {journal} {\bibinfo  {journal} {Phys. Rev. Lett.}\ }\textbf {\bibinfo
  {volume} {93}},\ \bibinfo {pages} {045901} (\bibinfo {year}
  {2004})}\BibitemShut {NoStop}%
\bibitem [{\citenamefont {Meschke}\ \emph {et~al.}(2006)\citenamefont
  {Meschke}, \citenamefont {Guichard},\ and\ \citenamefont
  {Pekola}}]{meschke-nature-2006}%
  \BibitemOpen
  \bibfield  {author} {\bibinfo {author} {\bibfnamefont {M.}~\bibnamefont
  {Meschke}}, \bibinfo {author} {\bibfnamefont {W.}~\bibnamefont {Guichard}}, \
  and\ \bibinfo {author} {\bibfnamefont {J.~P.}\ \bibnamefont {Pekola}},\
  }\href {\doibase 10.1038/nature05276} {\bibfield  {journal} {\bibinfo
  {journal} {Nature (London)}\ }\textbf {\bibinfo {volume} {444}},\ \bibinfo
  {pages} {187} (\bibinfo {year} {2006})}\BibitemShut {NoStop}%
\bibitem [{\citenamefont {Jezouin}\ \emph {et~al.}(2013)\citenamefont
  {Jezouin}, \citenamefont {Parmentier}, \citenamefont {Anthore}, \citenamefont
  {Gennser}, \citenamefont {Cavanna}, \citenamefont {Jin},\ and\ \citenamefont
  {Pierre}}]{jezouin-science-2013}%
  \BibitemOpen
  \bibfield  {author} {\bibinfo {author} {\bibfnamefont {S.}~\bibnamefont
  {Jezouin}}, \bibinfo {author} {\bibfnamefont {F.~D.}\ \bibnamefont
  {Parmentier}}, \bibinfo {author} {\bibfnamefont {A.}~\bibnamefont {Anthore}},
  \bibinfo {author} {\bibfnamefont {U.}~\bibnamefont {Gennser}}, \bibinfo
  {author} {\bibfnamefont {A.}~\bibnamefont {Cavanna}}, \bibinfo {author}
  {\bibfnamefont {Y.}~\bibnamefont {Jin}}, \ and\ \bibinfo {author}
  {\bibfnamefont {F.}~\bibnamefont {Pierre}},\ }\href {\doibase
  10.1126/science.1241912} {\bibfield  {journal} {\bibinfo  {journal}
  {Science}\ }\textbf {\bibinfo {volume} {342}},\ \bibinfo {pages} {601}
  (\bibinfo {year} {2013})}\BibitemShut {NoStop}%
\bibitem [{\citenamefont {Pascal}\ \emph {et~al.}(2011)\citenamefont {Pascal},
  \citenamefont {Courtois},\ and\ \citenamefont {Hekking}}]{pascal-prb-2011}%
  \BibitemOpen
  \bibfield  {author} {\bibinfo {author} {\bibfnamefont {L.~M.~A.}\
  \bibnamefont {Pascal}}, \bibinfo {author} {\bibfnamefont {H.}~\bibnamefont
  {Courtois}}, \ and\ \bibinfo {author} {\bibfnamefont {F.~W.~J.}\ \bibnamefont
  {Hekking}},\ }\href {\doibase 10.1103/PhysRevB.83.125113} {\bibfield
  {journal} {\bibinfo  {journal} {Phys. Rev. B}\ }\textbf {\bibinfo {volume}
  {83}},\ \bibinfo {pages} {125113} (\bibinfo {year} {2011})}\BibitemShut
  {NoStop}%
\bibitem [{\citenamefont {Okamoto}\ and\ \citenamefont
  {Massalski}(1985)}]{okamoto-1985}%
  \BibitemOpen
  \bibfield  {author} {\bibinfo {author} {\bibfnamefont {H.}~\bibnamefont
  {Okamoto}}\ and\ \bibinfo {author} {\bibfnamefont {T.~B.}\ \bibnamefont
  {Massalski}},\ }\href {\doibase 10.1007/bf02880404} {\bibfield  {journal}
  {\bibinfo  {journal} {Bull. All. Ph. Diagr.}\ }\textbf {\bibinfo {volume}
  {6}},\ \bibinfo {pages} {229} (\bibinfo {year} {1985})}\BibitemShut {NoStop}%
\bibitem [{\citenamefont {Peltonen}\ \emph {et~al.}(2010)\citenamefont
  {Peltonen}, \citenamefont {Virtanen}, \citenamefont {Meschke}, \citenamefont
  {Koski}, \citenamefont {Heikkil\"a},\ and\ \citenamefont
  {Pekola}}]{peltonen-prl-2010}%
  \BibitemOpen
  \bibfield  {author} {\bibinfo {author} {\bibfnamefont {J.~T.}\ \bibnamefont
  {Peltonen}}, \bibinfo {author} {\bibfnamefont {P.}~\bibnamefont {Virtanen}},
  \bibinfo {author} {\bibfnamefont {M.}~\bibnamefont {Meschke}}, \bibinfo
  {author} {\bibfnamefont {J.~V.}\ \bibnamefont {Koski}}, \bibinfo {author}
  {\bibfnamefont {T.~T.}\ \bibnamefont {Heikkil\"a}}, \ and\ \bibinfo {author}
  {\bibfnamefont {J.~P.}\ \bibnamefont {Pekola}},\ }\href {\doibase
  10.1103/PhysRevLett.105.097004} {\bibfield  {journal} {\bibinfo  {journal}
  {Phys. Rev. Lett.}\ }\textbf {\bibinfo {volume} {105}},\ \bibinfo {pages}
  {097004} (\bibinfo {year} {2010})}\BibitemShut {NoStop}%
\bibitem [{\citenamefont {Bezuglyi}\ and\ \citenamefont
  {Vinokur}(2003)}]{bezuglyi-prl-2003}%
  \BibitemOpen
  \bibfield  {author} {\bibinfo {author} {\bibfnamefont {E.~V.}\ \bibnamefont
  {Bezuglyi}}\ and\ \bibinfo {author} {\bibfnamefont {V.}~\bibnamefont
  {Vinokur}},\ }\href {\doibase 10.1103/physrevlett.91.137002} {\bibfield
  {journal} {\bibinfo  {journal} {Phys. Rev. Lett.}\ }\textbf {\bibinfo
  {volume} {91}},\ \bibinfo {pages} {137002} (\bibinfo {year}
  {2003})}\BibitemShut {NoStop}%
\bibitem [{\citenamefont {Dubos}\ \emph {et~al.}(2001)\citenamefont {Dubos},
  \citenamefont {Courtois}, \citenamefont {Pannetier}, \citenamefont {Wilhelm},
  \citenamefont {Zaikin},\ and\ \citenamefont {Sch\"{o}n}}]{dubos-prb-2001}%
  \BibitemOpen
  \bibfield  {author} {\bibinfo {author} {\bibfnamefont {P.}~\bibnamefont
  {Dubos}}, \bibinfo {author} {\bibfnamefont {H.}~\bibnamefont {Courtois}},
  \bibinfo {author} {\bibfnamefont {B.}~\bibnamefont {Pannetier}}, \bibinfo
  {author} {\bibfnamefont {F.~K.}\ \bibnamefont {Wilhelm}}, \bibinfo {author}
  {\bibfnamefont {A.~D.}\ \bibnamefont {Zaikin}}, \ and\ \bibinfo {author}
  {\bibfnamefont {G.}~\bibnamefont {Sch\"{o}n}},\ }\href {\doibase
  10.1103/physrevb.63.064502} {\bibfield  {journal} {\bibinfo  {journal} {Phys.
  Rev. B}\ }\textbf {\bibinfo {volume} {63}},\ \bibinfo {pages} {064502}
  (\bibinfo {year} {2001})}\BibitemShut {NoStop}%
\bibitem [{\citenamefont {Giazotto}\ \emph {et~al.}(2006)\citenamefont
  {Giazotto}, \citenamefont {Heikkil\"a}, \citenamefont {Luukanen},
  \citenamefont {Savin},\ and\ \citenamefont {Pekola}}]{giazotto-rmp-2006}%
  \BibitemOpen
  \bibfield  {author} {\bibinfo {author} {\bibfnamefont {F.}~\bibnamefont
  {Giazotto}}, \bibinfo {author} {\bibfnamefont {T.~T.}\ \bibnamefont
  {Heikkil\"a}}, \bibinfo {author} {\bibfnamefont {A.}~\bibnamefont
  {Luukanen}}, \bibinfo {author} {\bibfnamefont {A.~M.}\ \bibnamefont {Savin}},
  \ and\ \bibinfo {author} {\bibfnamefont {J.~P.}\ \bibnamefont {Pekola}},\
  }\href {\doibase 10.1103/RevModPhys.78.217} {\bibfield  {journal} {\bibinfo
  {journal} {Rev. Mod. Phys.}\ }\textbf {\bibinfo {volume} {78}},\ \bibinfo
  {pages} {217} (\bibinfo {year} {2006})}\BibitemShut {NoStop}%
\bibitem [{\citenamefont {Meschke}\ \emph {et~al.}(2009)\citenamefont
  {Meschke}, \citenamefont {Peltonen}, \citenamefont {Courtois},\ and\
  \citenamefont {Pekola}}]{meschke-jltp-2009}%
  \BibitemOpen
  \bibfield  {author} {\bibinfo {author} {\bibfnamefont {M.}~\bibnamefont
  {Meschke}}, \bibinfo {author} {\bibfnamefont {J.~T.}\ \bibnamefont
  {Peltonen}}, \bibinfo {author} {\bibfnamefont {H.}~\bibnamefont {Courtois}},
  \ and\ \bibinfo {author} {\bibfnamefont {J.~P.}\ \bibnamefont {Pekola}},\
  }\href {\doibase 10.1007/s10909-008-9854-y} {\bibfield  {journal} {\bibinfo
  {journal} {J. Low Temp. Phys.}\ }\textbf {\bibinfo {volume} {154}},\ \bibinfo
  {pages} {190} (\bibinfo {year} {2009})}\BibitemShut {NoStop}%
\bibitem [{\citenamefont {Virtanen}\ \emph {et~al.}(2011)\citenamefont
  {Virtanen}, \citenamefont {Bergeret}, \citenamefont {Cuevas},\ and\
  \citenamefont {Heikkil\"{a}}}]{virtanen-prb-2011}%
  \BibitemOpen
  \bibfield  {author} {\bibinfo {author} {\bibfnamefont {P.}~\bibnamefont
  {Virtanen}}, \bibinfo {author} {\bibfnamefont {F.~S.}\ \bibnamefont
  {Bergeret}}, \bibinfo {author} {\bibfnamefont {J.~C.}\ \bibnamefont
  {Cuevas}}, \ and\ \bibinfo {author} {\bibfnamefont {T.~T.}\ \bibnamefont
  {Heikkil\"{a}}},\ }\href {\doibase 10.1103/physrevb.83.144514} {\bibfield
  {journal} {\bibinfo  {journal} {Phys. Rev. B}\ }\textbf {\bibinfo {volume}
  {83}},\ \bibinfo {pages} {144514} (\bibinfo {year} {2011})}\BibitemShut
  {NoStop}%
\bibitem [{\citenamefont {Courtois}\ \emph {et~al.}(2008)\citenamefont
  {Courtois}, \citenamefont {Meschke}, \citenamefont {Peltonen},\ and\
  \citenamefont {Pekola}}]{courtois-prl-2008}%
  \BibitemOpen
  \bibfield  {author} {\bibinfo {author} {\bibfnamefont {H.}~\bibnamefont
  {Courtois}}, \bibinfo {author} {\bibfnamefont {M.}~\bibnamefont {Meschke}},
  \bibinfo {author} {\bibfnamefont {J.~T.}\ \bibnamefont {Peltonen}}, \ and\
  \bibinfo {author} {\bibfnamefont {J.~P.}\ \bibnamefont {Pekola}},\ }\href
  {\doibase 10.1103/physrevlett.101.067002} {\bibfield  {journal} {\bibinfo
  {journal} {Phys. Rev. Lett.}\ }\textbf {\bibinfo {volume} {101}},\ \bibinfo
  {pages} {067002} (\bibinfo {year} {2008})}\BibitemShut {NoStop}%
\bibitem [{\citenamefont {G\"{o}ppl}\ \emph {et~al.}(2008)\citenamefont
  {G\"{o}ppl}, \citenamefont {Fragner}, \citenamefont {Baur}, \citenamefont
  {Bianchetti}, \citenamefont {Filipp}, \citenamefont {Fink}, \citenamefont
  {Leek}, \citenamefont {Puebla}, \citenamefont {Steffen},\ and\ \citenamefont
  {Wallraff}}]{goppl-jap-2008}%
  \BibitemOpen
  \bibfield  {author} {\bibinfo {author} {\bibfnamefont {M.}~\bibnamefont
  {G\"{o}ppl}}, \bibinfo {author} {\bibfnamefont {A.}~\bibnamefont {Fragner}},
  \bibinfo {author} {\bibfnamefont {M.}~\bibnamefont {Baur}}, \bibinfo {author}
  {\bibfnamefont {R.}~\bibnamefont {Bianchetti}}, \bibinfo {author}
  {\bibfnamefont {S.}~\bibnamefont {Filipp}}, \bibinfo {author} {\bibfnamefont
  {J.~M.}\ \bibnamefont {Fink}}, \bibinfo {author} {\bibfnamefont {P.~J.}\
  \bibnamefont {Leek}}, \bibinfo {author} {\bibfnamefont {G.}~\bibnamefont
  {Puebla}}, \bibinfo {author} {\bibfnamefont {L.}~\bibnamefont {Steffen}}, \
  and\ \bibinfo {author} {\bibfnamefont {A.}~\bibnamefont {Wallraff}},\ }\href
  {\doibase 10.1063/1.3010859} {\bibfield  {journal} {\bibinfo  {journal} {J.
  Appl. Phys.}\ }\textbf {\bibinfo {volume} {104}},\ \bibinfo {pages} {113904}
  (\bibinfo {year} {2008})}\BibitemShut {NoStop}%
\bibitem [{\citenamefont {Dassonneville}\ \emph {et~al.}(2013)\citenamefont
  {Dassonneville}, \citenamefont {Ferrier}, \citenamefont {Gu\'{e}ron},\ and\
  \citenamefont {Bouchiat}}]{dassonneville-prl-2013}%
  \BibitemOpen
  \bibfield  {author} {\bibinfo {author} {\bibfnamefont {B.}~\bibnamefont
  {Dassonneville}}, \bibinfo {author} {\bibfnamefont {M.}~\bibnamefont
  {Ferrier}}, \bibinfo {author} {\bibfnamefont {S.}~\bibnamefont {Gu\'{e}ron}},
  \ and\ \bibinfo {author} {\bibfnamefont {H.}~\bibnamefont {Bouchiat}},\
  }\href {\doibase 10.1103/physrevlett.110.217001} {\bibfield  {journal}
  {\bibinfo  {journal} {Phys. Rev. Lett.}\ }\textbf {\bibinfo {volume} {110}},\
  \bibinfo {pages} {217001} (\bibinfo {year} {2013})}\BibitemShut {NoStop}%
\bibitem [{\citenamefont {Richards}(1994)}]{richards-jap-1994}%
  \BibitemOpen
  \bibfield  {author} {\bibinfo {author} {\bibfnamefont {P.~L.}\ \bibnamefont
  {Richards}},\ }\href {\doibase 10.1063/1.357128} {\bibfield  {journal}
  {\bibinfo  {journal} {J. Appl. Phys.}\ }\textbf {\bibinfo {volume} {76}},\
  \bibinfo {pages} {1} (\bibinfo {year} {1994})}\BibitemShut {NoStop}%
\bibitem [{\citenamefont {Gershenzon}\ \emph {et~al.}(1988)\citenamefont
  {Gershenzon}, \citenamefont {Gol’tsman}, \citenamefont {Elant’ev},
  \citenamefont {Karasik},\ and\ \citenamefont
  {Potoskuev}}]{gershenzon-sjltp-1988}%
  \BibitemOpen
  \bibfield  {author} {\bibinfo {author} {\bibfnamefont {E.~M.}\ \bibnamefont
  {Gershenzon}}, \bibinfo {author} {\bibfnamefont {G.~N.}\ \bibnamefont
  {Gol’tsman}}, \bibinfo {author} {\bibfnamefont {A.~I.}\ \bibnamefont
  {Elant’ev}}, \bibinfo {author} {\bibfnamefont {B.~S.}\ \bibnamefont
  {Karasik}}, \ and\ \bibinfo {author} {\bibfnamefont {S.~E.}\ \bibnamefont
  {Potoskuev}},\ }\href@noop {} {\bibfield  {journal} {\bibinfo  {journal}
  {Sov. J. Low Temp. Phys.}\ }\textbf {\bibinfo {volume} {14}},\ \bibinfo
  {pages} {753} (\bibinfo {year} {1988})}\BibitemShut {NoStop}%
\bibitem [{\citenamefont {Elantev}\ and\ \citenamefont
  {Karasik}(1989)}]{elantev-sjltp-1989}%
  \BibitemOpen
  \bibfield  {author} {\bibinfo {author} {\bibfnamefont {A.~I.}\ \bibnamefont
  {Elantev}}\ and\ \bibinfo {author} {\bibfnamefont {B.~S.}\ \bibnamefont
  {Karasik}},\ }\href@noop {} {\bibfield  {journal} {\bibinfo  {journal} {Sov.
  J. Low Temp. Phys.}\ }\textbf {\bibinfo {volume} {15}},\ \bibinfo {pages}
  {379} (\bibinfo {year} {1989})}\BibitemShut {NoStop}%
\bibitem [{\citenamefont {Karasik}\ \emph {et~al.}(2009)\citenamefont
  {Karasik}, \citenamefont {Pereverzev}, \citenamefont {Olaya}, \citenamefont
  {Wei}, \citenamefont {Gershenson},\ and\ \citenamefont
  {Sergeev}}]{karasik-ieee-2009}%
  \BibitemOpen
  \bibfield  {author} {\bibinfo {author} {\bibfnamefont {B.~S.}\ \bibnamefont
  {Karasik}}, \bibinfo {author} {\bibfnamefont {S.~V.}\ \bibnamefont
  {Pereverzev}}, \bibinfo {author} {\bibfnamefont {D.}~\bibnamefont {Olaya}},
  \bibinfo {author} {\bibfnamefont {J.}~\bibnamefont {Wei}}, \bibinfo {author}
  {\bibfnamefont {M.~E.}\ \bibnamefont {Gershenson}}, \ and\ \bibinfo {author}
  {\bibfnamefont {A.~V.}\ \bibnamefont {Sergeev}},\ }\href {\doibase
  10.1109/tasc.2009.2019426} {\bibfield  {journal} {\bibinfo  {journal} {IEEE
  Trans. Appl. Supercond.}\ }\textbf {\bibinfo {volume} {19}},\ \bibinfo
  {pages} {532} (\bibinfo {year} {2009})}\BibitemShut {NoStop}%
\bibitem [{\citenamefont {Muhonen}\ \emph {et~al.}(2012)\citenamefont
  {Muhonen}, \citenamefont {Meschke},\ and\ \citenamefont
  {Pekola}}]{muhonen-rpp-2012}%
  \BibitemOpen
  \bibfield  {author} {\bibinfo {author} {\bibfnamefont {J.~T.}\ \bibnamefont
  {Muhonen}}, \bibinfo {author} {\bibfnamefont {M.}~\bibnamefont {Meschke}}, \
  and\ \bibinfo {author} {\bibfnamefont {J.~P.}\ \bibnamefont {Pekola}},\
  }\href {\doibase 10.1088/0034-4885/75/4/046501} {\bibfield  {journal}
  {\bibinfo  {journal} {Rep. Prog. Phys.}\ }\textbf {\bibinfo {volume} {75}},\
  \bibinfo {pages} {046501} (\bibinfo {year} {2012})}\BibitemShut {NoStop}%
\bibitem [{\citenamefont {Chandrasekhar}(2009)}]{chandrasekhar-sst-2009}%
  \BibitemOpen
  \bibfield  {author} {\bibinfo {author} {\bibfnamefont {V.}~\bibnamefont
  {Chandrasekhar}},\ }\href {\doibase 10.1088/0953-2048/22/8/083001} {\bibfield
   {journal} {\bibinfo  {journal} {Supercond. Sci. Technol.}\ }\textbf
  {\bibinfo {volume} {22}},\ \bibinfo {pages} {083001} (\bibinfo {year}
  {2009})}\BibitemShut {NoStop}%
\bibitem [{\citenamefont {Arutyunov}\ \emph {et~al.}(2008)\citenamefont
  {Arutyunov}, \citenamefont {Golubev},\ and\ \citenamefont
  {Zaikin}}]{arutyunov-2008}%
  \BibitemOpen
  \bibfield  {author} {\bibinfo {author} {\bibfnamefont {K.~Y.}\ \bibnamefont
  {Arutyunov}}, \bibinfo {author} {\bibfnamefont {D.~S.}\ \bibnamefont
  {Golubev}}, \ and\ \bibinfo {author} {\bibfnamefont {A.~D.}\ \bibnamefont
  {Zaikin}},\ }\href {\doibase 10.1016/j.physrep.2008.04.009} {\bibfield
  {journal} {\bibinfo  {journal} {Phys. Rep.}\ }\textbf {\bibinfo {volume}
  {464}},\ \bibinfo {pages} {1} (\bibinfo {year} {2008})}\BibitemShut {NoStop}%
\bibitem [{\citenamefont {Timofeev}\ \emph {et~al.}(2009)\citenamefont
  {Timofeev}, \citenamefont {Helle}, \citenamefont {Meschke}, \citenamefont
  {M\"{o}tt\"{o}nen},\ and\ \citenamefont {Pekola}}]{timofeev-prl-2009b}%
  \BibitemOpen
  \bibfield  {author} {\bibinfo {author} {\bibfnamefont {A.~V.}\ \bibnamefont
  {Timofeev}}, \bibinfo {author} {\bibfnamefont {M.}~\bibnamefont {Helle}},
  \bibinfo {author} {\bibfnamefont {M.}~\bibnamefont {Meschke}}, \bibinfo
  {author} {\bibfnamefont {M.}~\bibnamefont {M\"{o}tt\"{o}nen}}, \ and\
  \bibinfo {author} {\bibfnamefont {J.~P.}\ \bibnamefont {Pekola}},\ }\href
  {\doibase 10.1103/physrevlett.102.200801} {\bibfield  {journal} {\bibinfo
  {journal} {Phys. Rev. Lett.}\ }\textbf {\bibinfo {volume} {102}},\ \bibinfo
  {pages} {200801} (\bibinfo {year} {2009})}\BibitemShut {NoStop}%
\bibitem [{\citenamefont {Muhonen}\ \emph {et~al.}(2009)\citenamefont
  {Muhonen}, \citenamefont {Niskanen}, \citenamefont {Meschke}, \citenamefont
  {Yu}, \citenamefont {Tsai}, \citenamefont {Sainiemi}, \citenamefont
  {Franssila},\ and\ \citenamefont {Pekola}}]{muhonen-apl-2009}%
  \BibitemOpen
  \bibfield  {author} {\bibinfo {author} {\bibfnamefont {J.~T.}\ \bibnamefont
  {Muhonen}}, \bibinfo {author} {\bibfnamefont {A.~O.}\ \bibnamefont
  {Niskanen}}, \bibinfo {author} {\bibfnamefont {M.}~\bibnamefont {Meschke}},
  \bibinfo {author} {\bibnamefont {Yu}}, \bibinfo {author} {\bibfnamefont
  {J.~S.}\ \bibnamefont {Tsai}}, \bibinfo {author} {\bibfnamefont
  {L.}~\bibnamefont {Sainiemi}}, \bibinfo {author} {\bibfnamefont
  {S.}~\bibnamefont {Franssila}}, \ and\ \bibinfo {author} {\bibfnamefont
  {J.~P.}\ \bibnamefont {Pekola}},\ }\href {\doibase 10.1063/1.3080668}
  {\bibfield  {journal} {\bibinfo  {journal} {Appl. Phys. Lett.}\ }\textbf
  {\bibinfo {volume} {94}},\ \bibinfo {pages} {073101} (\bibinfo {year}
  {2009})}\BibitemShut {NoStop}%
\bibitem [{\citenamefont {Bardeen}\ \emph {et~al.}(1959)\citenamefont
  {Bardeen}, \citenamefont {Rickayzen},\ and\ \citenamefont
  {Tewordt}}]{bardeen-pr-1959}%
  \BibitemOpen
  \bibfield  {author} {\bibinfo {author} {\bibfnamefont {J.}~\bibnamefont
  {Bardeen}}, \bibinfo {author} {\bibfnamefont {G.}~\bibnamefont {Rickayzen}},
  \ and\ \bibinfo {author} {\bibfnamefont {L.}~\bibnamefont {Tewordt}},\ }\href
  {\doibase 10.1103/PhysRev.113.982} {\bibfield  {journal} {\bibinfo  {journal}
  {Phys. Rev.}\ }\textbf {\bibinfo {volume} {113}},\ \bibinfo {pages} {982}
  (\bibinfo {year} {1959})}\BibitemShut {NoStop}%
\bibitem [{\citenamefont {Aumentado}\ \emph {et~al.}(2004)\citenamefont
  {Aumentado}, \citenamefont {Keller}, \citenamefont {Martinis},\ and\
  \citenamefont {Devoret}}]{aumentado-prl-2004}%
  \BibitemOpen
  \bibfield  {author} {\bibinfo {author} {\bibfnamefont {J.}~\bibnamefont
  {Aumentado}}, \bibinfo {author} {\bibfnamefont {M.~W.}\ \bibnamefont
  {Keller}}, \bibinfo {author} {\bibfnamefont {J.~M.}\ \bibnamefont
  {Martinis}}, \ and\ \bibinfo {author} {\bibfnamefont {M.~H.}\ \bibnamefont
  {Devoret}},\ }\href {\doibase 10.1103/physrevlett.92.066802} {\bibfield
  {journal} {\bibinfo  {journal} {Phys. Rev. Lett.}\ }\textbf {\bibinfo
  {volume} {92}},\ \bibinfo {pages} {066802} (\bibinfo {year}
  {2004})}\BibitemShut {NoStop}%
\bibitem [{\citenamefont {de~Visser}\ \emph {et~al.}(2011)\citenamefont
  {de~Visser}, \citenamefont {Baselmans}, \citenamefont {Diener}, \citenamefont
  {Yates}, \citenamefont {Endo},\ and\ \citenamefont
  {Klapwijk}}]{devisser-prl-2011}%
  \BibitemOpen
  \bibfield  {author} {\bibinfo {author} {\bibfnamefont {P.~J.}\ \bibnamefont
  {de~Visser}}, \bibinfo {author} {\bibfnamefont {J.~J.~A.}\ \bibnamefont
  {Baselmans}}, \bibinfo {author} {\bibfnamefont {P.}~\bibnamefont {Diener}},
  \bibinfo {author} {\bibfnamefont {S.~J.~C.}\ \bibnamefont {Yates}}, \bibinfo
  {author} {\bibfnamefont {A.}~\bibnamefont {Endo}}, \ and\ \bibinfo {author}
  {\bibfnamefont {T.~M.}\ \bibnamefont {Klapwijk}},\ }\href {\doibase
  10.1103/physrevlett.106.167004} {\bibfield  {journal} {\bibinfo  {journal}
  {Phys. Rev. Lett.}\ }\textbf {\bibinfo {volume} {106}},\ \bibinfo {pages}
  {167004} (\bibinfo {year} {2011})}\BibitemShut {NoStop}%
\bibitem [{\citenamefont {Bezuglyi}\ \emph {et~al.}(2001)\citenamefont
  {Bezuglyi}, \citenamefont {Bratus'}, \citenamefont {Shumeiko},\ and\
  \citenamefont {Wendin}}]{bezuglyi-prb-2001}%
  \BibitemOpen
  \bibfield  {author} {\bibinfo {author} {\bibfnamefont {E.~V.}\ \bibnamefont
  {Bezuglyi}}, \bibinfo {author} {\bibfnamefont {E.~N.}\ \bibnamefont
  {Bratus'}}, \bibinfo {author} {\bibfnamefont {V.~S.}\ \bibnamefont
  {Shumeiko}}, \ and\ \bibinfo {author} {\bibfnamefont {G.}~\bibnamefont
  {Wendin}},\ }\href {\doibase 10.1103/physrevb.63.100501} {\bibfield
  {journal} {\bibinfo  {journal} {Phys. Rev. B}\ }\textbf {\bibinfo {volume}
  {63}},\ \bibinfo {pages} {100501} (\bibinfo {year} {2001})}\BibitemShut
  {NoStop}%
\bibitem [{\citenamefont {Pascal}\ \emph {et~al.}(2013)\citenamefont {Pascal},
  \citenamefont {Fay}, \citenamefont {Winkelmann},\ and\ \citenamefont
  {Courtois}}]{pascal-prb-2013}%
  \BibitemOpen
  \bibfield  {author} {\bibinfo {author} {\bibfnamefont {L.~M.~A.}\
  \bibnamefont {Pascal}}, \bibinfo {author} {\bibfnamefont {A.}~\bibnamefont
  {Fay}}, \bibinfo {author} {\bibfnamefont {C.~B.}\ \bibnamefont {Winkelmann}},
  \ and\ \bibinfo {author} {\bibfnamefont {H.}~\bibnamefont {Courtois}},\
  }\href {\doibase 10.1103/PhysRevB.88.100502} {\bibfield  {journal} {\bibinfo
  {journal} {Phys. Rev. B}\ }\textbf {\bibinfo {volume} {88}},\ \bibinfo
  {pages} {100502} (\bibinfo {year} {2013})}\BibitemShut {NoStop}%
\end{thebibliography}%

\end{document}